\documentclass[journal=apchd5,manuscript=article]{achemso}
\usepackage[T1]{fontenc}
\usepackage{amssymb}
\usepackage{xfrac}
\usepackage{geometry}
\usepackage[version=3]{mhchem} 

\author{Matej Bubaš}
\affiliation{Ruđer Bošković Institute, Bijenička cesta 54, Zagreb 10000, Croatia}

\alsoaffiliation{National Institute of Chemistry, Hajdrihova 19, Ljubljana 1000, Slovenia}
\author{Jordi Sancho-Parramon}
\email{jsancho@irb.hr}
\affiliation
{Ruđer Bošković Institute, Bijenička cesta 54, Zagreb 10000, Croatia}

\title[Maximizing nanoparticle light absorption]
  {Maximizing nanoparticle light absorption: \\ size, geometry, and a prospect for metal alloys}

\keywords{absorption, plasmonics, alloys, nanoparticles, retardation, radiation damping, dynamic depolarization}

\begin{document}







\begin{abstract}
  In this work we show how to maximize absorption of plasmonic nanoparticles in terms of size, geometry and material. For that reason the interaction of nanoparticles with light was decomposed into different effects. We determined that the main effect dictating the optimal amount of optical losses is radiation damping, and how it depends on nanoparticle size and geometry. Based on this, we find that for many combinations of sizes and geometries losses in pure metals are far from optimal. To overcome the aforementioned issue, alloying is presented as straightforward and flexible way of modulating the optical losses. Furthermore, strategies for tuning the optical losses to values above, between, and even below those in pure plasmonic metals are developed in terms of selecting the right alloy composition. In some cases, alloys showed a multifold increase in absorption when compared to pure plasmonic metals. The physical reasons governing such changes are elucidated based on the electronic structure changes during alloying of different metals, which enables generalization of the results to other systems. Besides increasing absorption, electronic structure changes can also be utilized for channeling the absorbed energy to suit different purposes, such as hot carrier generation for photocatalysis or solar energy harvesting. Overall, these results establish alloying as a powerful tool for designing nanostructures for applications that utilize light absorption.

\end{abstract}

\section{Introduction}

Plasmonics plays a central role in photonics due to the unique properties of surface waves excited at metal-dielectric interfaces \cite{maier2007plasmonics}. The expansion of plasmonics towards real-world applications has been traditionally characterized by efforts to alleviate the negative effects of optical losses, such as field-enhancement quenching or propagation length reduction \cite{boltasseva2011low,khurgin2015deal}. In contrast, current trends propose a paradigm shift and prioritize applications where the benefits of losses may outweigh the drawbacks \cite{boriskina2017losses}, focusing on solar energy harvesting \cite{cushing2016progress}, super-absorbers \cite{lei2018ultra},  photothermal effects \cite{yang2022thermoplasmonics} or plasmonic photocatalysis \cite{zhang2013plasmonic}.

The early stages of plasmonics were also characterized by the design of devices based on geometry and morphology optimization, using noble metals and other low-loss alternatives as the only materials of choice \cite{west2010searching}. Nowadays, material engineering has become an additional degree of freedom in the design of nanostructures, with alloying having a leading part \cite{gutierrez2020plasmonics}.  Metal alloying enables tuning the optical response between those of the alloyed materials \cite{gong2018band} being particularly advantageous for hot carrier devices, absorbers, catalysis and photovoltaics \cite{gong2020emergent, tiburski2022engineering, fojt2024}. Moreover, alloying unlocks emergent properties not present in pure metals, such as low-energy electronic transitions that can enhance infrared absorption \cite{bubaš2021dft} and open new pathways for tailoring hot carrier energy distribution \cite{bubas2024hotcarrier}.  

Here we put forward the optical losses tuning provided by alloying as a compelling pathway to optimize absorption in plasmonic nanostructures. For the sake of clarity, we note that the optical losses discussed here are given by the imaginary part of the dielectric function - i.e. optical losses include all the dissipative mechanisms in the material that result with absorption of light in the nanoparticles. 

First, the electromagnetic response of dipole-like particles is revisited to illustrate how finite-size effects determine the amount of losses that maximize the absorption cross-section. It turns out that optical losses are especially beneficial in particles with a large aspect ratio. Afterwards, we describe how band structure engineering enabled by alloying can be used to increase nanoparticle absorption beyond the values attainable using pure metals. In a thorough discussion, alloying is presented as an effective mechanism for achieving optimal plasmonic response in absorption-based applications such as photocatalysis, photovoltaics, and photothermal cancer therapy. 

\section{Optimal losses in plasmonic particles}

Optical losses incorporate all absorptive loss mechanisms, including several connected to plasmon decay \cite{boriskina2017losses, brown2016plasmondecay}, and not just Ohmic losses which result in heating. The utilizations of different types of optical losses are discussed later in the text, in the Outlook section. Optical losses are commonly considered detrimental for absorption and sometimes the Rayleigh approximation is invoked to support that view 
\cite{bohren1983can,fan2014light,yezekyan2020maximizing}.
However, such picture is oversimplified and can even lead to the paradoxical conclusion that non-absorbing particles maximize the absorption efficiency (Supplementary Information, Section 1). 

\begin{figure*}[ht!]
\centering
\includegraphics[width=\linewidth]{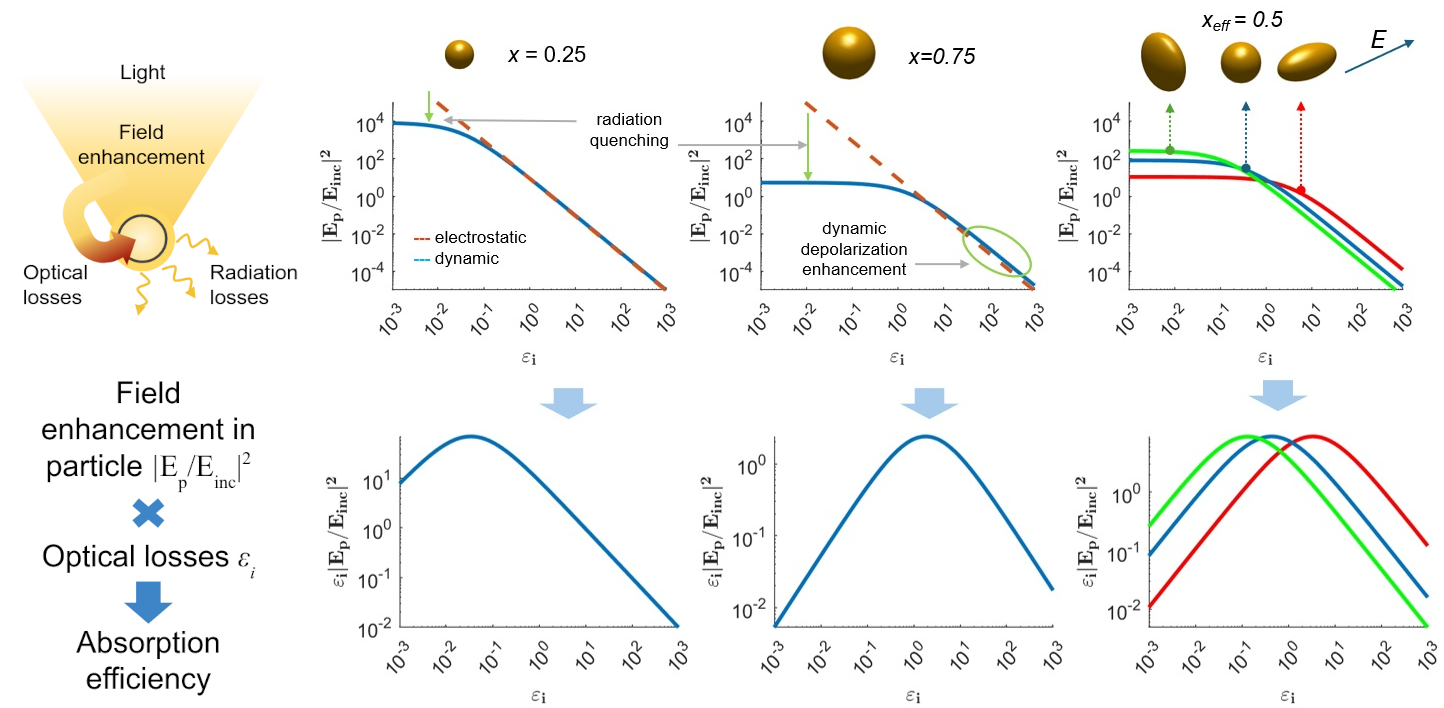}
\caption{The absorption efficiency ($Q_{abs}$) of a small particle is proportional to the product of the intensity enhancement inside the particle and the imaginary part of the dielectric function ($\varepsilon_i$). The electrostatic approximation predicts that the field intensity enhancement inside the particle diverges for $\varepsilon_i \rightarrow 0$ (dashed red lines). When retardation corrections are included (Section 1 in Supplementary Information), the field enhancement of a small particle is quenched by radiation losses and remains finite (left column, top row). As a consequence, $Q_{abs}$  is maximized for small (yet non-zero) values of $\varepsilon_i$ (left column, bottom row). For larger particles, quenching by radiation losses dominates over a wider range of $\varepsilon_i$, and $Q_{abs}$ becomes maximum at larger $\varepsilon_i$ values. Additionally, dynamic depolarization partially quenches the effect of losses (middle column). In the case of non-spherical shapes depolarization effects are more evident:  prolate (oblate) particles polarized along the symmetry axis require larger (lower) $\varepsilon_i$ values to maximize absorption (right column).} 
\label{ref:figENH}
\end{figure*}

Several studies have addressed the question of the optimal amount of losses that maximize absorption using more nuanced schemes such as analyzing the exact Mie theory or using impedance matching concepts \cite{tribelsky2011anomalous,tretyakov2014maximizing,wu2011large}. In this section, we revisit this topic from a more intuitive perspective that clarifies the role of size and geometry in determining the optimal losses in nanoparticles. Assuming a particle of radius $a$ and complex dielectric function $\varepsilon_p = \varepsilon_r+ i \varepsilon_i$ embedded in a medium with dielectric function $\varepsilon_m$ and excited by a plane wave with wavelength $\lambda$, the absorption efficiency $Q_{abs}$ is calculated through the integration of losses over the particle volume:

\begin{equation}
    Q_{abs} = \frac{1}{\pi a^2} \frac{1}{2 I_{inc}} \int dV \varepsilon_i \varepsilon_0 \omega \lvert \textbf{E}_{p}\rvert^2   \approxeq \frac{4x}{3\varepsilon_m} \varepsilon_i \left|\frac{\textbf{E}_p}{\textbf{E}_{inc}}\right|^2 ,
    \label{eq:ohmic}
\end{equation}
with $I_{inc}$ the intensity of the incoming plane wave, $\varepsilon_0$  the vacuum permittivity and $x = 2 \pi \sqrt{\varepsilon_m}a/\lambda$ is the particle size factor. The right side of Equation \ref{eq:ohmic} is obtained assuming that the field inside the particle $\left(\textbf{E}_p\right) $ is constant.  The absorption efficiency then becomes the product between particle losses ($\varepsilon_i$) and the field intensity enhancement inside the particle, which is quenched by $\varepsilon_i$. In the electrostatic limit, the Rayleigh approximation is recovered (see Supplementary Information), corresponding to the situation in which the field enhancement increases with no bound for $\varepsilon_i \rightarrow 0$ (Figure \ref{ref:figENH}, dashed lines in top row).

However, the field inside a particle always remains finite because it is also dampened by radiation losses, which are proportional to the particle volume \cite{draine1988discrete}. Thus, radiation losses are the dominant field-quenching mechanism when optical losses are small, i.e. the field inside the particle is independent of $\varepsilon_i$ when $\varepsilon_i \rightarrow 0$ (Figure \ref{ref:figENH}, top row). Therefore, the largest absorption efficiency is not obtained for $\varepsilon_i \rightarrow 0$, as the apparent paradox stemming from using Rayleigh approximation would suggest. Instead it is close to the largest $\varepsilon_i$ that is \textit{small enough} to prevent that optical losses become the dominating field quenching mechanism (Figure \ref{ref:figENH}, bottom row). 

It therefore turns out that, to properly address the problem of absorption maximization in nanoparticles, radiation losses need to be taken into account for any particle size \cite{tribelsky2011anomalous, tretyakov2014maximizing}. By including retardation and radiative damping corrections in the particle polarizability or, equivalently, in the field enhancement inside the particle (see Supplementary Information), the maximum absorption efficiency is obtained for:

\begin{equation}
    \frac{\varepsilon_{i,max}}{\varepsilon_m} \approxeq 2 x^3 + \frac{14}{5} x^5+\mathcal{O}\left({x^7}\right) ,
    \label{eq:epsimax}
\end{equation}
that provides a very good approximation to the result obtained using the Mie theory up to relatively large values of the size parameter ($x \lesssim 1$, Figure S1). Now $\varepsilon_i \rightarrow 0$ maximizes the absorption efficiency only when $x \rightarrow 0$, i.e. in the point-like limit. Note, however, that the total amount of absorbed light, given by the absorption cross section ($C_{abs} = \pi a^2 Q_{abs}$), remains always finite. Its maximum value is nearly independent of the particle size and scales with $\lambda^2$ (see \cite{tribelsky2011anomalous,tretyakov2014maximizing} and Supplementary Information). 

\subsection{Influence of nanoparticle size on absorption efficiency}
As the particle size increases, radiation losses become more prominent. This means that they represent the main field quenching mechanism over a broader range of $\varepsilon_i$. As stated earlier, the absorption efficiency is maximized when optical losses overtake radiation losses as the dominant quenching mechanism. Therefore, for larger particles the optical losses, reflected in the values of $\varepsilon_i$, also need to be larger to maximize the absorption efficiency (Figure \ref{ref:figENH}, middle column). 

The need for higher optical losses is further magnified due to the increasing significance of the higher order terms on the right side of Equation \ref{eq:epsimax} with increasing size factor. The influence of higher order terms  shifts the condition for maximum absorption to larger $\varepsilon_i$ values. This shift occurs due to the dynamic depolarization of the particle, i.e. the effect of dephasing in the field radiated by different particle locations that build up the depolarization field in the particle. It is shown in the Supplementary information that dynamic depolarization, accounted for in higher order terms, partially alleviates the quenching of the electromagnetic field caused by optical losses \cite{meier1983enhanced}.

\subsection{Influence of particle geometry on absorption efficiency}
Extending the analysis to geometry of the particle, it can be seen that, in comparison to spherical particle, higher losses are desirable for elongated particles, while for oblate (shortened on one axis) lower losses maximize absorption. Again, as was the case for the influence of particle size, the optimal amount of optical losses is governed by the amount of radiation losses for a given geometry. 

To grasp the effect of particle geometry on absorption efficiency, let us consider the case of spheroidal particles. First assume that the incoming light is polarized along the symmetry axis of a prolate spheroid, i.e., along its longer semi-axis (inset in Figure \ref{ref:figENH}, top row, right column). By taking into account the polarizability of a spheroidal particle corrected for retardation and radiative damping \cite{meier1983enhanced,moroz2009depolarization}, the maximum absorption efficiency is obtained for (see Supplementary Information):
\begin{equation}
    \frac{\varepsilon_{i,max}}{\varepsilon_m} \approxeq  \frac{2}{9 L^2} x_{eff}^3 +  \frac{4 D R^{2/3}}{27L^3}x_{eff}^5 + \mathcal{O}{\left(x_{eff}^7\right)} ,
    \label{eq:eiprolate}
\end{equation}
with $L$ being the static depolarization factor, $D$ the dynamic depolarization factor, $R$ the ratio between the particle short and long semi-axes and $x_{eff}$ an effective size parameter. When $R$ is reduced, $L$ decreases and $D$ grows \cite{moroz2009depolarization}.

Thus, at a fixed volume, more elongated particles require higher losses to maximize $Q_{abs}$ (Figure \ref{ref:figENH}, right column, red lines). This trend can be understood by noting that the radiative damping term ($-\sfrac{2i}{9}(\varepsilon_p-\varepsilon_m)  x_{eff}^3$, see Supplementary Information) depends on the particle volume but not directly on its shape. However, the prefactor $(\varepsilon_p-\varepsilon_m)$ changes with the particle shape, since the plasmon resonance shifts to more negative $\varepsilon_r$ for prolate particles polarized along its symmetry axis \cite{maier2007plasmonics}. In that case, quenching by radiative damping becomes stronger, shifting the condition for maximum absorption to higher $\varepsilon_i$. Finally, the effects of dynamic depolarization, represented in the $x^5$ term in Eq. \ref{eq:eiprolate}, are more pronounced than in the case of a spherical particle, which further increases the preference for higher optical losses when maximizing absorption. 

The opposite behavior is observed for an oblate particle polarized along its symmetry axis (shorter semi-axis): as the particle shape departs from sphericity, lower losses are required to maximize absorption (Figure \ref{ref:figENH}, green lines). In this case the shift of the plasmon resonance to less negative values reduces the magnitude of $(\varepsilon_p-\varepsilon_m)$ and weakens radiation damping, shifting the condition for maximum absorption to lower $\varepsilon_i$ values. This reasoning can be extended to the case of polarization along non-symmetry directions, for it is based only on the magnitude of $\varepsilon_p$ at the plasmon resonance. Thus, regardless of the particle symmetry, maximization of absorption for light polarized along a major (or minor) axis requires higher (or lower) losses than in the case of a spherical particle with the same volume, as supported by the results of numerical simulations (Figure S3).

\section{Metal alloys for optimal absorption}
The discussion up to this point illustrates how absorption maximization for a given combination of size and geometry requires an optimal amount of losses. The issue in achieving the maximum absorption is that the optical loss values provided by pure materials are often far from optimal. To overcome this issue, the tunability of optical losses provided by alloying can be remarkably advantageous. In the following sections we provide strategies for absorption maximization in three possible cases: by tuning the losses between, above, and below those exhibited by pure constitutent metals. For each case we explain the tunability by providing physical insights based on electronic structure changes, which also allows us to generalize the findings to other alloyed systems. 

\subsection{Alloying to obtain intermediate losses} 

\begin{figure*}[h!]
\centering
\includegraphics[width=\linewidth]{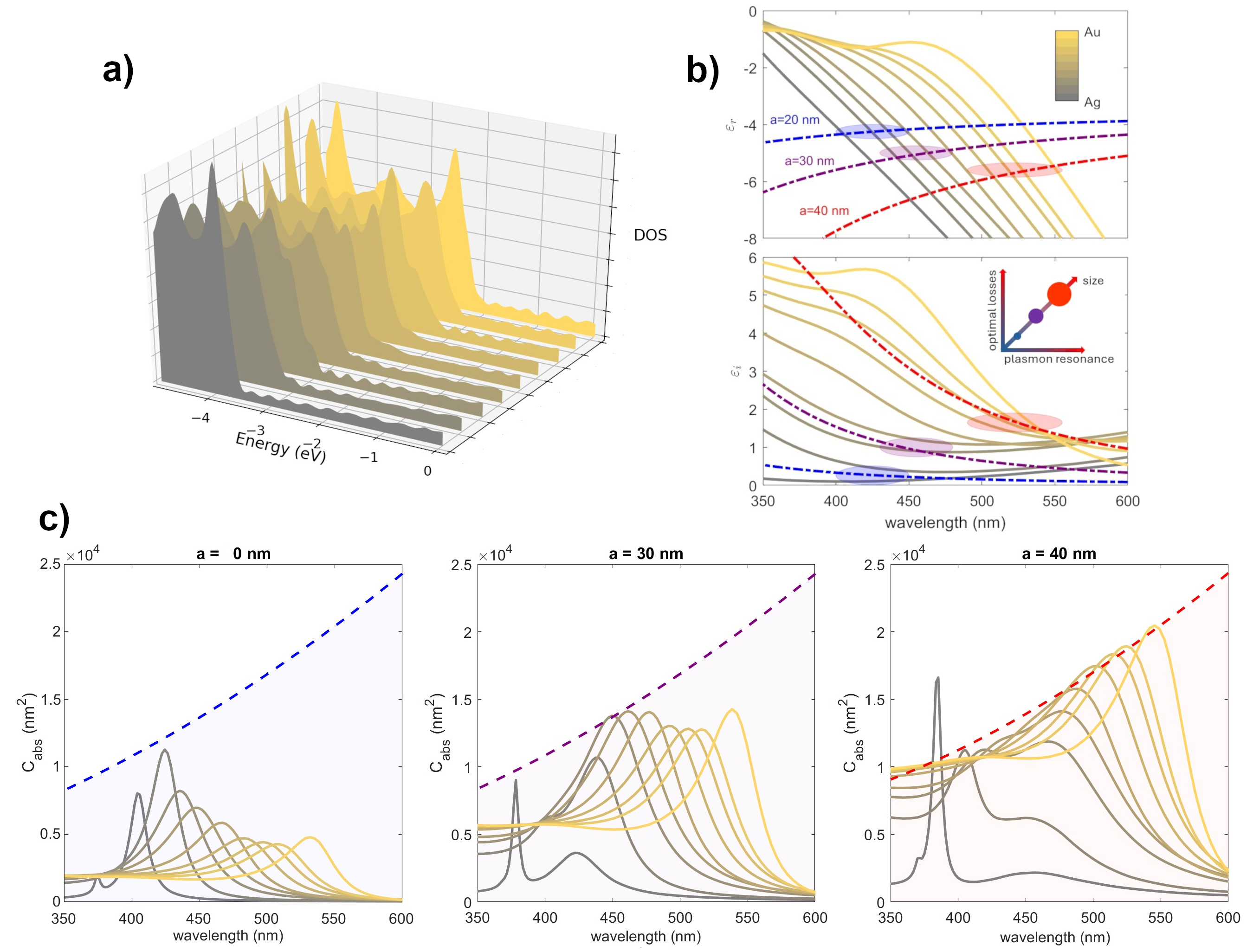}
\caption{a) Density of states (DOS) for Au (yellow), Ag (grey), and their alloys, calculated by DFT. The region with a high DOS corresponds to the d-band. The d-band edge, that in large part determines the optical properties of these metals, is gradually shifting to higher values from Ag to Au. Fermi level is set to 0 eV. b) Real (top) and imaginary (bottom) parts of the dielectric function of Au (yellow), Ag (grey) and their alloys (grey/yellow). Dashed lines indicate the values of the $\varepsilon_r$ and $\varepsilon_i$ that maximize absorption for particles of 20 (blue), 30 (purple) and 40 (red) nm embedded in a water-like environment ($\varepsilon_m=1.77$). The coloured areas indicate the composition and wavelength regions in which both the plasmon resonance and the optimal absorption conditions are approximately satisfied. The dielectric functions of Au, Ag and their alloys were determined by ellipsometry.
c) Absorption cross sections for particles of Ag, Au and their alloys with radius equal to 20 (left), 30 (middle) and 40 (right) nm in a water-like environment computed using the Mie theory. Dashed lines: theoretically maximum attainable absorption cross section computed with the Mie theory accounting for the electric dipole response only. Note that computed data can overpass this limit due to the influence of quadrupole or higher multipole orders.
\cite{pena2014optical}.}
\label{ref:figepsAgAu}
\end{figure*}

A suitable system for analysis and illustration of this tuning strategy is a Au-Ag alloy system, combining the two most prevalent metals in plasmonics research \cite{gong2018band}. Both metals possess similar electronic configurations, characterized by fully filled low-energy d-bands, and a similar plasma frequency \cite{johnson1972optical}. As alloy composition varies, the d-band edge gradually shifts between limiting value of pure metals (Figure \ref{ref:figepsAgAu}a). With it, the onset of interband transitions starting from the d-bands also shifts, which is reflected in the gradual changes in the dielectric function (Figure \ref{ref:figepsAgAu}b). This gradual change allows for precise tuning, a quality particularly beneficial for meeting the conditions for maximal absorption.

To maximize absorption, two conditions have to be met simultaneously: 

i) Nanoparticles have to be excited near or at plasmon resonance, which results in strong field enhancement and is determined by $\varepsilon_r$. For particle sizes up to a few tens of nanometers, Ag, Au and their binary alloys satisfy the resonance condition in the visible range (Equation S5, dashed lines in Figure \ref{ref:figepsAgAu}b, top).  

ii) Their optical losses, reflected in the imaginary part of the dielectric function $\varepsilon_i$, have to be near optimal values for absorption at that wavelength range (Equation \ref{eq:epsimax}, dashed lines in Figure \ref{ref:figepsAgAu}b, bottom). 

Meeting both conditions can be attainable only within a restricted range of compositions that depends on the particle size. This is best seen in the spectral dependence of the absorption cross sections for different particle sizes (Figure \ref{ref:figepsAgAu}c). For small nanoparticles, characterized by a small $x$, low losses are required. For such particles Ag is the best choice since, in the range of interest, photons do not have the required energy to excite transitions that involve the d-band. However, as particle size increases, the optimal composition shifts to more lossy Au-rich alloys with increasing contribution of d-band related interband transitions. Note that although absorption can increase with higher losses, scattering and extinction always decrease due to quenching of the field enhancement. Thus, for these two quantities Ag always performs better than alloys or Au (Figure S4). This interplay can be used to tune the absorption-to-scattering ratio through material selection \cite{langhammer2007absorption}, instead of geometric optimization.

\begin{figure}[ht!]
\centering
\includegraphics[width=0.6\linewidth]{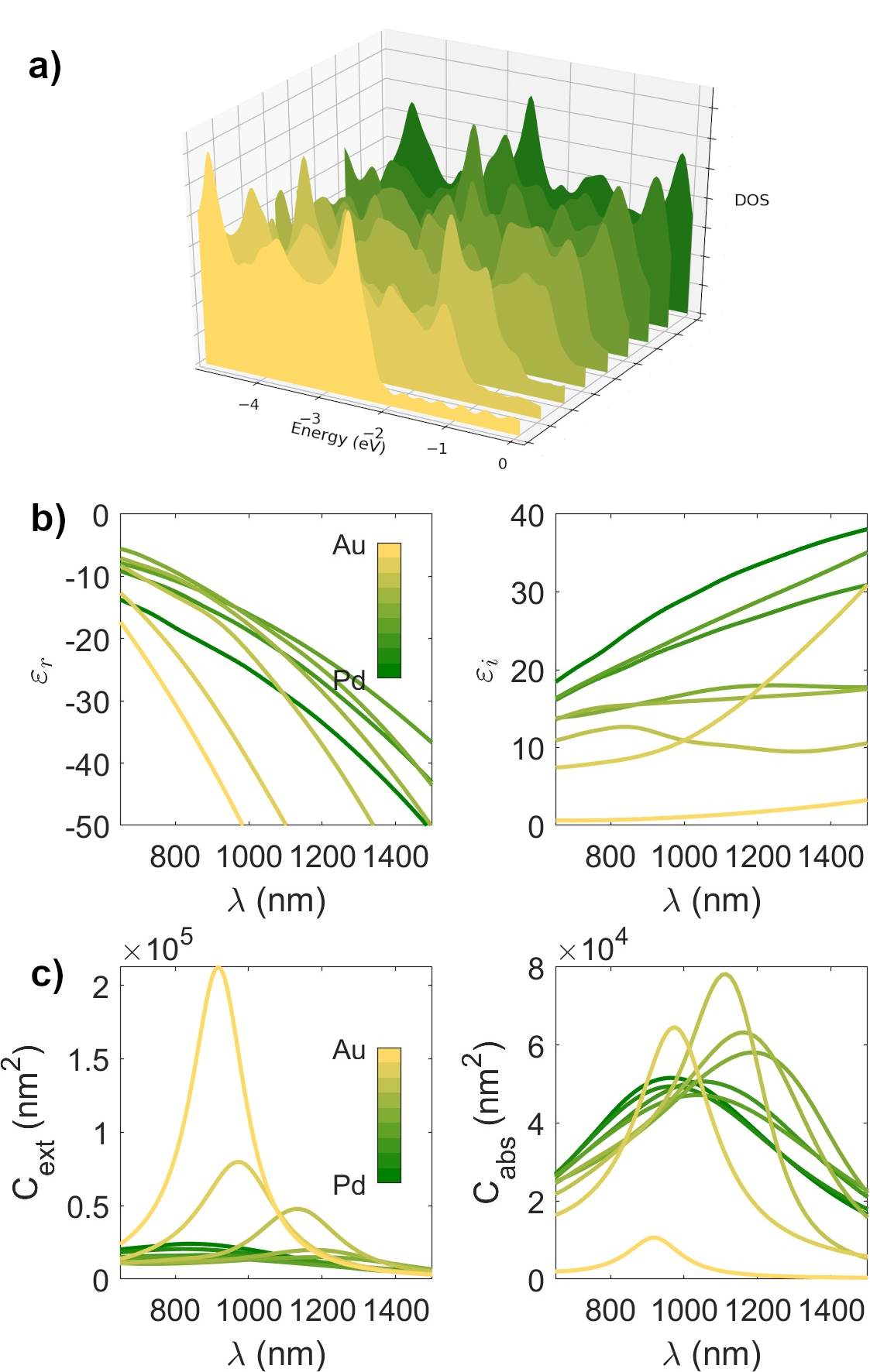}
\caption{a) Density of states (DOS) for Au (yellow), Pd (green), and their alloys, calculated by DFT. A region with a high DOS corresponds to the d-band. The d-band edge abruptly rises in energy as Pd is alloyed with Au, and continues to rise more slowly, finally crossing the Fermi level (set to 0 eV) at a large Pd ratio. b) Real (left) and imaginary (right) part of the dielectric function of AuPd alloys, calculated by density functional theory as reported in \cite{bubaš2021dft}. Note that optical losses $\varepsilon_i$ rise abruptly as a small ratio of Pd is alloyed with Au. c) Extinction (left) and absorption (right) cross-sections of nanorods  made of different PdAu alloys with  200 nm length and 60 nm diameter and having hemispherical caps. The nanorod is assumed to be in a water-like medium and excited with light polarized along its long axis. Although extinction favors minimal values of $\varepsilon_i$, maximum of absorption is achieved with intermediate values of $\varepsilon_i$}
\label{figPdAu}
\end{figure}

The second system of interest is a Pd-Au alloy, a mixture of increasing interest in plasmonics due to its photocatalytic potential \cite{liu2016design} and enhanced hot carrier generation by low energy photons \cite{stofela2020noble}. Figure \ref{figPdAu}a shows that, in contrast to Au, Pd has higher energy d-states which form shallow d-bands. As can be seen in the figure, the Pd d-band edge even crosses the Fermi level, leading to filled and unfilled d-states very close in energy. Due to a high density of states available for low-energy transitions, Pd displays much larger losses than Au in the near-infrared range (Figure \ref{figPdAu}b, right panel). This expands the limits for the tuning range, but also results in more abrupt changes in the dielectric function due to a rapid shift in the d-band energy upon alloying \cite{bubaš2021dft, bubas2024phd, rahm2020library}.  

Although the tuning might not be as fine as for Au-Ag alloy system, the advantage of a much wider tuning range achieved by incorporating a metal such as Pd, that is characterized by high losses, can be illustrated by considering a case of a nanorod (200 nm length, 60 nm diameter) excited by light polarized along its long axis. The moderate aspect ratio of such a particle leads to a plasmon resonance in the near-infrared range (Figure \ref{figPdAu}c). Photon energy in the near-infrared range is below the interband threshold for Au where electrons in the d-states can be excited, but also above the range where the Drude term dominates and causes high losses. Thus, Au possesses very low losses in that region. The absorption characteristics are drastically improved by increasing the losses, with the best absorption shown by alloys that contain a moderate amount of Pd. In such systems the absorption cross-section shows a large increase, up to 800\% when compared to pure Au. 

\subsection{Alloying to obtain larger losses}
To increase absorption by alloying, Au-Ag alloy can again be used, this time exploiting the emergent properties of alloying. In this case, the wavelength region of interest is below the energy necessary to excite electrons in d-bands. Namely, we focus on the high energy part of the near-infrared and lower energy part of the visible spectrum. It has been mentioned in previous section that Au has very low losses in that region, and Ag provides even lower losses. Interestingly, in the same region, optical losses of their alloys do not fall between the values of pure metals but instead smoothly increase above values of both metals when their ratio approaches 50\% (Figure S5). The origin of this behaviour is discussed at the end of this section. 

\begin{figure*}[ht!]
\centering
\includegraphics[width=0.26\textwidth]{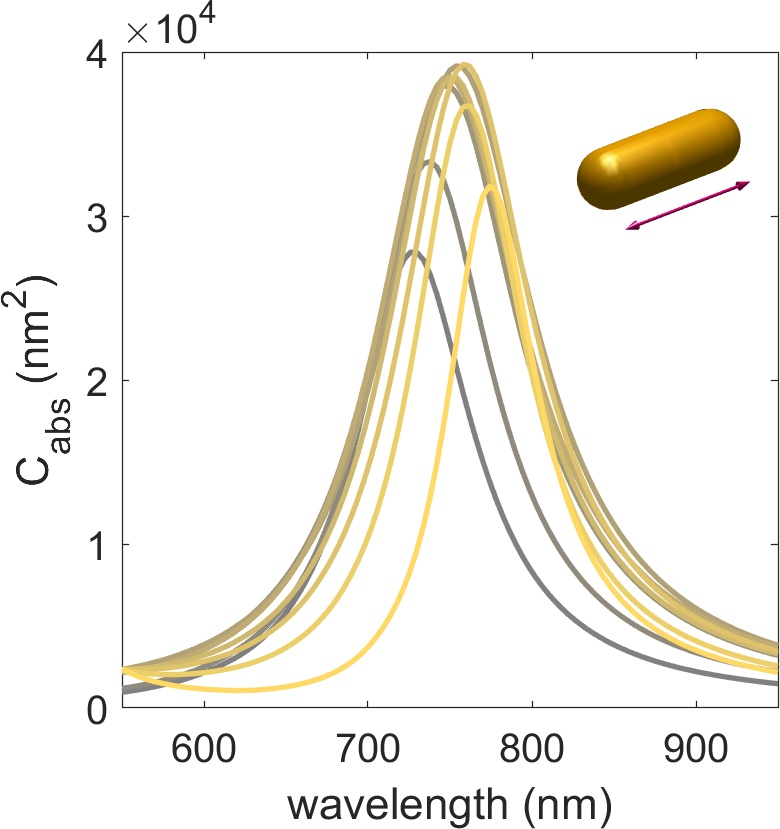}
\hspace{15pt}
\includegraphics[width=0.27\textwidth]{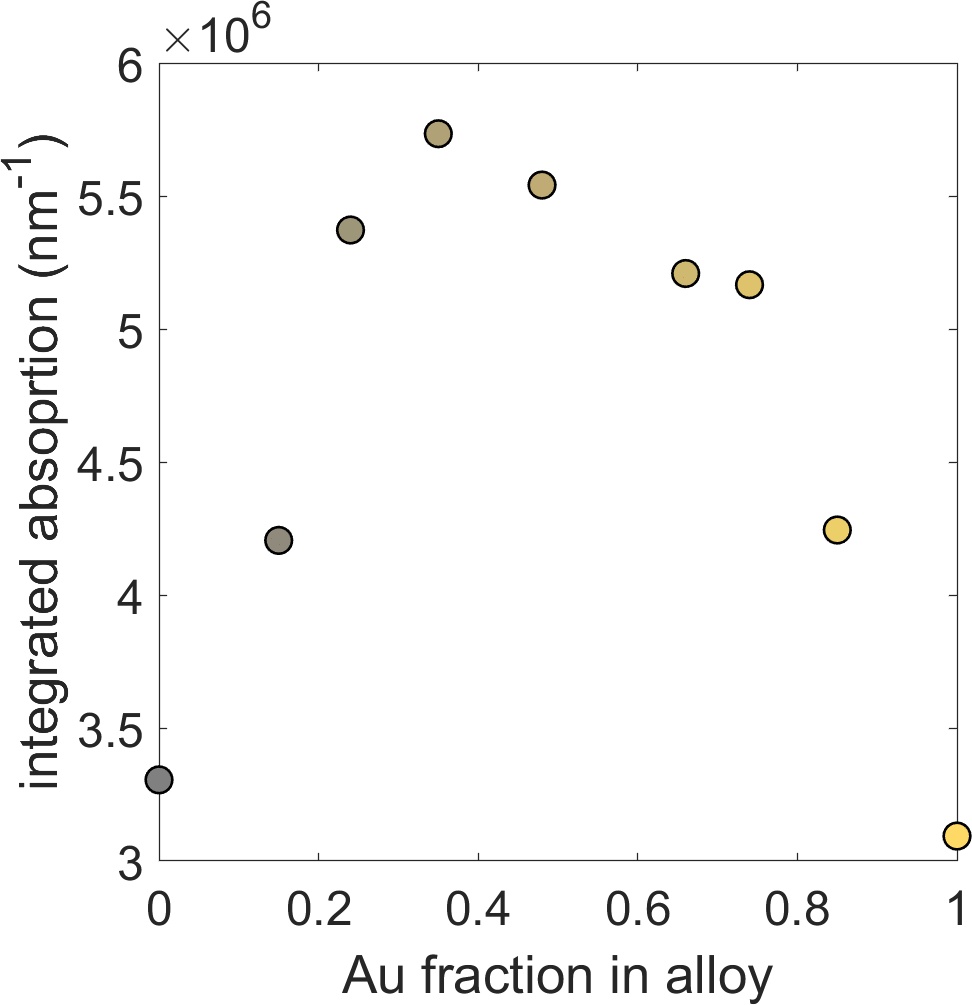}
\hspace{15pt}
\includegraphics[width=0.30\textwidth]{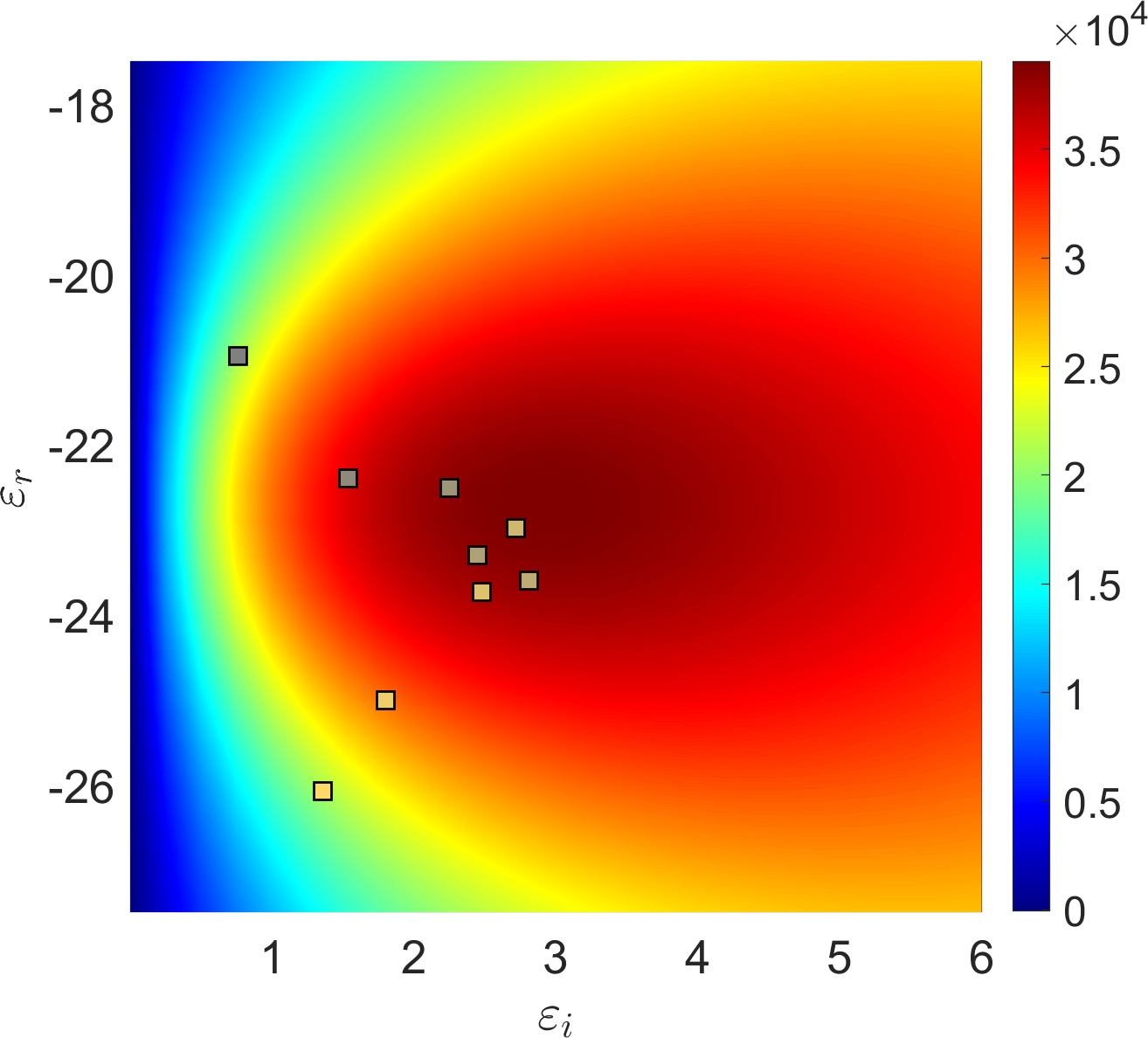}

\includegraphics[width=0.27\textwidth]{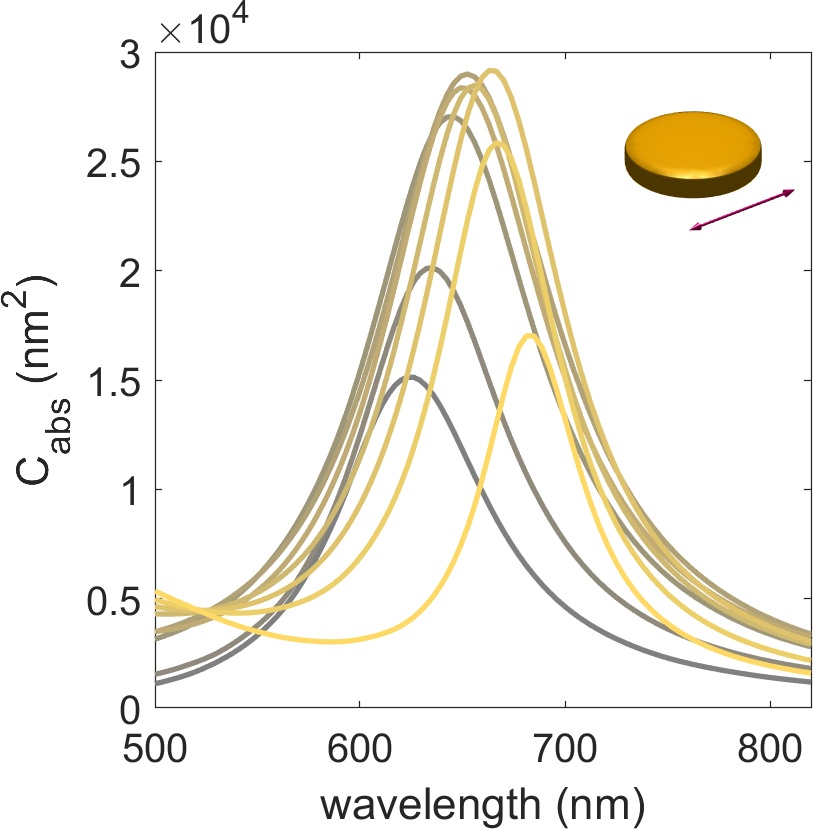}
\hspace{15pt}
\includegraphics[width=0.27\textwidth]{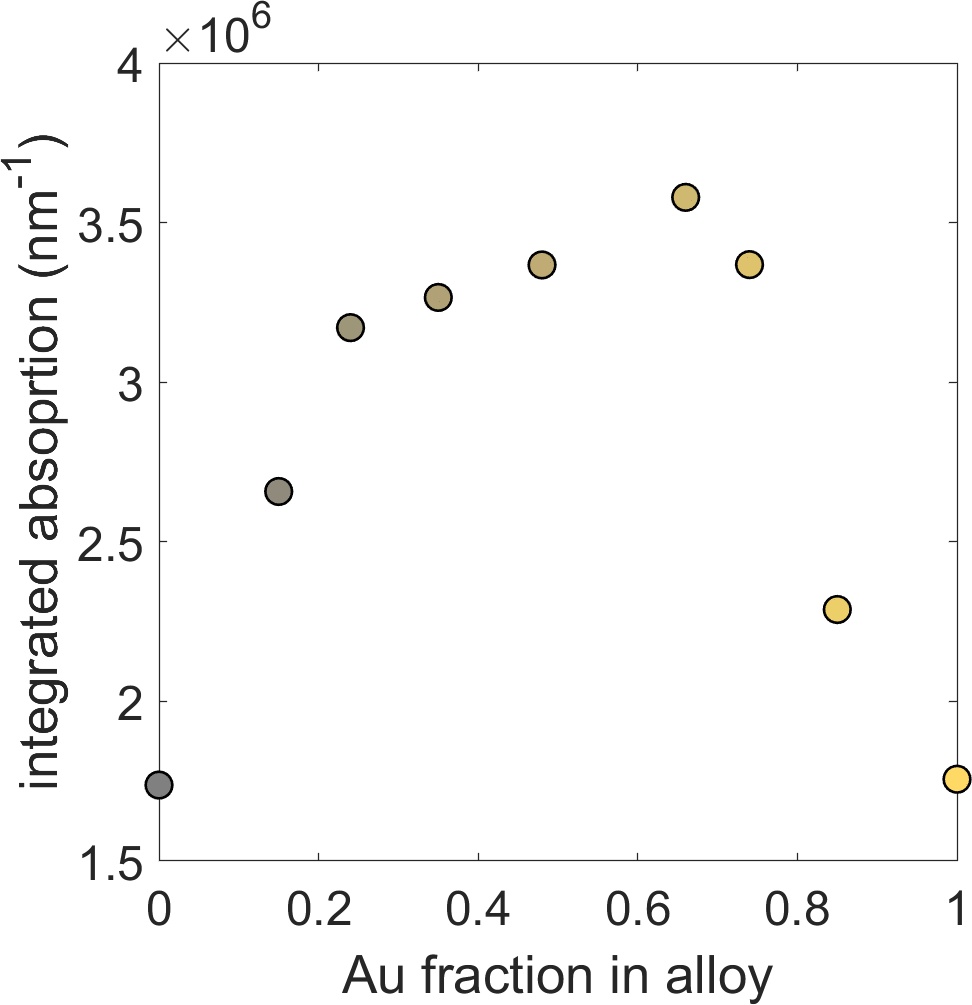}
\hspace{15pt}
\includegraphics[width=0.30\textwidth]{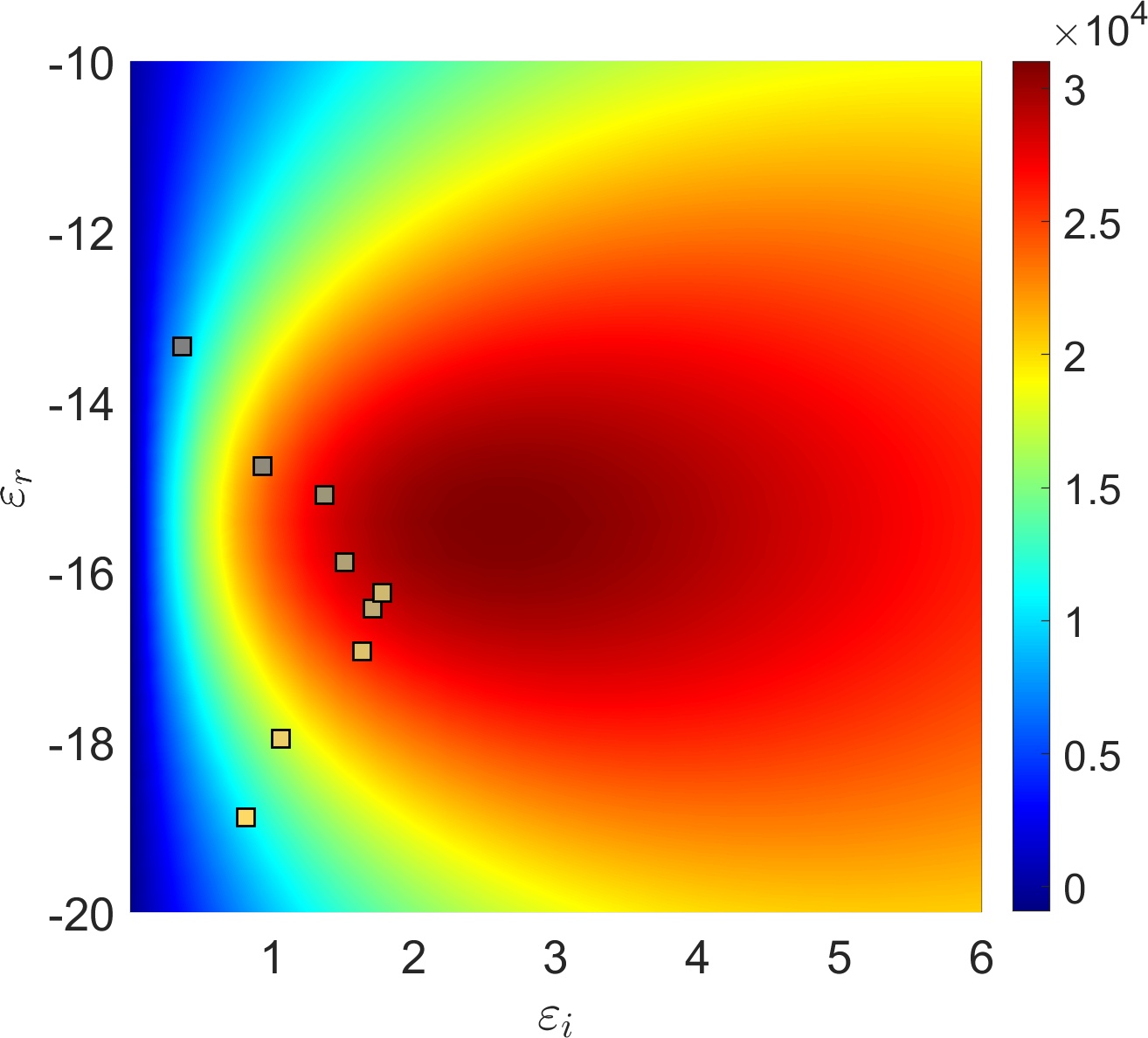}
\caption{Top: Absorption efficiency spectra of a nanorod (120 nm length, 40 nm diameter) of Ag-Au alloys alloy embedded in water polarized along its long axis (left), integrated absorption (middle) and map of the absorption efficiency as a function of $\varepsilon_r$ and $\varepsilon_i$ for the specific geometry. The values of the optical constants of alloys and pure metals are represented as dots. Bottom: same as top row, for a disk (100 nm diameter, 20 nm length) polarized along the disk diameter.}
\label{figAbsMapRod}
\end{figure*}

The enhanced infrared absorption can be beneficial when pure metals exhibit optical losses that are too low for maximizing absorption. This property is particularly useful for elongated particles, as previously described. We consider the case of a nanorod excited by light polarized along its long axis (Figure \ref{figAbsMapRod}). In this scenario, alloys display maximum absorption efficiency values up to 50\% higher than those of pure metals. Moreover, the broadening of the absorption peak caused by the additional optical losses in alloys enhances the advantage when considering total absorption in the displayed spectral range, reflected by the integrated absorption. When the absorption efficiency is integrated over the spectral range the increase for alloys rises up to up to 200\% when compared with pure metals. Since the plasmonic properties of nanorods are usually heavily dominated by scattering, their utilizations are primarily making use of that property. In contrast, the presented results expand the promising utilizations of alloy nanorods to absorption-based applications, such as light harvesting. Moreover, by using the fine tuning to shift optical losses above values presented by pure metals, the alloy nanorods can achieve absorption efficiency values near the theoretical maximum for that specific shape (right column map in Figure \ref{figAbsMapRod}). This finding can be extended to some other geometries, such as nanodisk excited along the disk diameter (Figure \ref{figAbsMapRod} bottom row). As determined in the Section 3, since the particle is polarized along the major axis, radiation losses are large and more optical losses will be required to match it and maximize absorption. Due to this fact, nanodisks also display significantly larger absorption efficiency values for alloys compared to pure metals, and even larger increase in integrated absorption. This suggests that, as for nanorods,  alloying expands their utilization towards uses that benefit from absorption. 

This behavior of the dielectric function and the optical losses is usually associated with conduction electrons and represented by a Drude-like contribution. It is experimentally observed that alloys exhibit larger infrared losses than their constituent materials (Figure S5), which is traditionally explained in terms of enhanced electronic scattering due to crystal lattice imperfections \cite{pena2014optical}. On the other hand, it has been recently shown that band folding and splitting, inherent to alloying, enables interband transitions in the conduction band(s) which require low energy photons (below the nominal interband threshold in pure metals) \cite{bubaš2021dft, bubas2024phd}. In fact, using computational methods and considering only the losses related to interband transitions unlocked by alloying, with other loss sources removed, the trend of loss increase with Au-Ag alloying has been qualitatively reproduced \cite{bubas2024phd}. These results suggest that the reason for the increased absorption in the range of interest are the emergent interband transitions in alloys of good plasmonic metals (those with low d-bands). Since the effect of loss increase in that region was observed for all combinations of low d-band metals, the results can be generalized to other alloys with analogous electronic structures. 

\subsection{Alloying to obtain lower losses}
\begin{figure}[h!]
\centering
\includegraphics[width=0.65\linewidth]{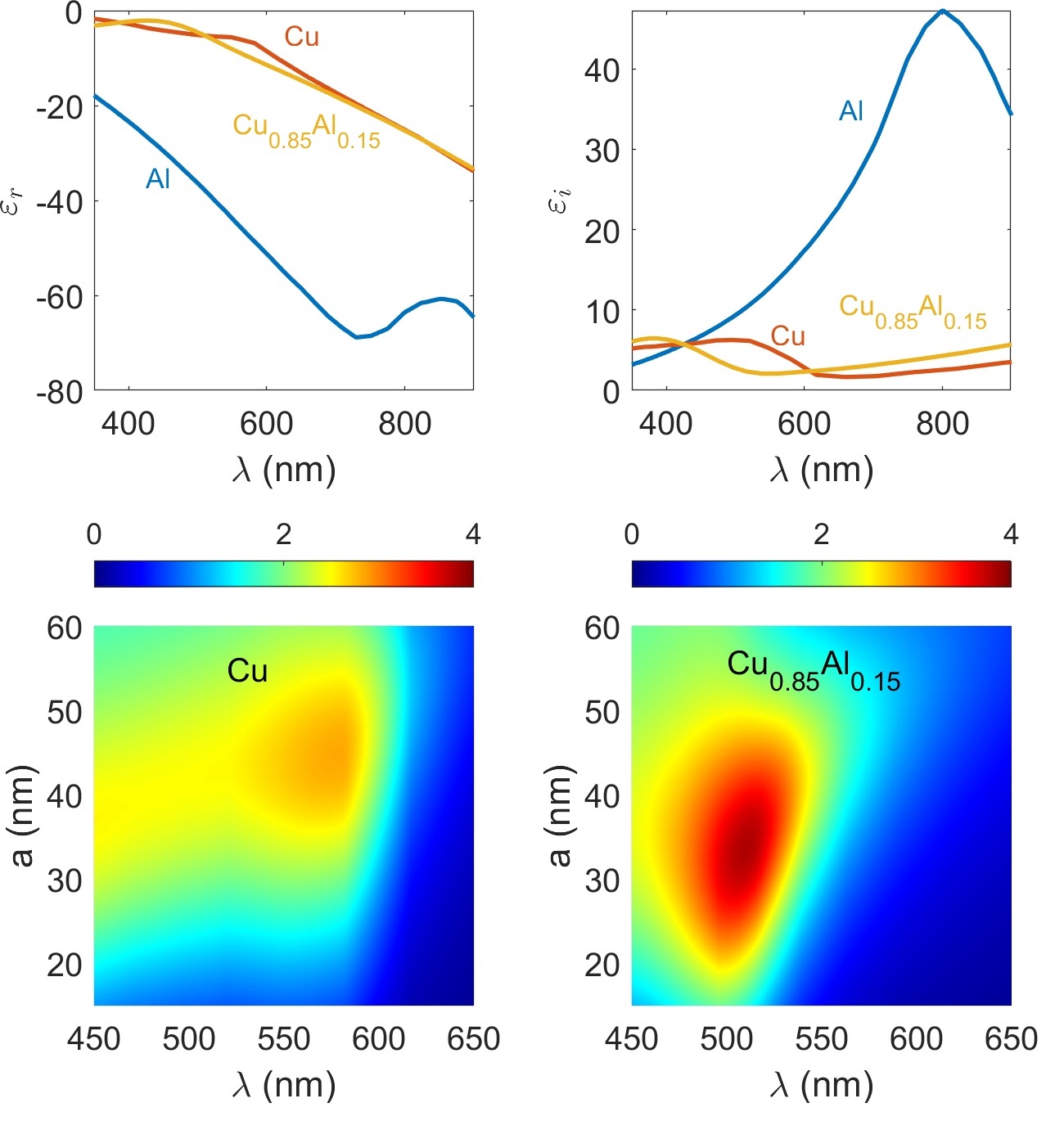}
\caption{Top: Real (left) and imaginary (part) of the dielectric function of Cu, Al and Cu$_{0.85}$Al$_{0.15}$ taken from literature (\cite{johnson1972optical}, \cite{palik1998handbook} and \cite{shahcheraghi2016anomalously} respectively). Bottom: absorption efficiency as a function of wavelength and particle radius for Cu (left) and Cu$_{0.85}$Al$_{0.15}$ spherical particles embedded in a water-like environment.}
\label{figCuAl}
\end{figure}

In this section, we complete the picture of optimizing absorption, by showing how to benefit from minimization of optical losses by alloying. To showcase loss minimization with respect to pure elements the analysis is extended to beyond just d-block elements to include Al, a p-shell element in an alloy with Cu. Typically for good plasmonic metals, Cu has an electronic structure with fully filled d-bands, similar to Ag and Au but with a slightly shallower d-band edge, resulting in interband transitions (onset around 600 nm) that quench its plasmonic response in most of the visible range. The addition of a moderate amount of Al ($\approx$ 15\%) to Cu mainly leads to lowering of the d-band edge \cite{shahcheraghi2016anomalously} and a corresponding interband transitions onset shift towards shorter wavelengths. Since the addition of Al does not significantly affect the real part of the dielectric function of Cu, this essentially results in reducing the optical losses that quench the plasmonic response of Cu in the visible range. The alloy losses are also lower when compared to Al, since Al possesses interband transitions in the entire energy range, especially at longer wavelengths ($\approx$ 800 nm) which lead to significant optical losses all the way to UV. However, when the amount of Al in the alloy is low, that property is not reflected in the optical losses of the alloy. Thus, in the spectral range between 450 and 600 nm, the Cu$_{0.85}$Al$_{0.15}$ alloy presents lower optical losses than both Cu and Al, as shown in Figure \ref{figCuAl}.    

In terms of nanoparticle size to optimize absorption, Cu particles require relatively large sizes to obtain high absorption efficiencies, following Equation \ref{eq:epsimax} (Figure \ref{figCuAl}, bottom left). In comparison, considerably larger absorption efficiency values can be achieved with Cu-Al alloy particles, and that the optimal particle sizes are in this case smaller (Figure \ref{figCuAl}, bottom right). This exemplifies that alloying can be beneficial for absorption not only for larger and nonspherical particles, but also for nanoparticles that are small and spherical. 

\section{Outlook}
The findings presented insofar can be implemented to benefit an array of different fields such as photocatalysis, solar energy harvesting, phototermal cancer therapy and several more. The underlying principle is directing more energy, obtained due to increasing absorption, to the system. The way the energy is transferred to the system can be different, and is dependent on the types of nonradiative losses in the nanoparticle. While some optical losses immediately lead to heat, other first lead to creation of highly energetic electrons and holes called hot carriers \cite{boriskina2017losses, brown2016plasmondecay}. Based on that, we divide the prospective contribution of highly absorbing alloyed nanostructures.

\textbf{Enhancing hot carrier-based applications:}
Hot carriers can be injected into a molecule, initiating a reaction \cite{, zhu2024molecular, lee2022surface, christopher2017hot}, or into a semiconductor such as a photovoltaic, facilitating solar energy conversion to electric current \cite{liu2022dynamic, tang2020plasmonic}. One way to enhance the optical loss channel related to hot carrier generation is by tuning the d-bands of metals to enhance either interband or intraband transitions \cite{bubas2024phd, bubas2024hotcarrier, douglas2016atomistic, sanches2022plasmon}. Recently, potential benefits for hot carrier generation in Pd-Au alloys were highlighted \cite{bubas2024hotcarrier}. As a caveat, it was noted that introduction of an efficient loss channel might also reduce absorption due to field quenching. The implication was drawn that the compromise between increased efficiency and less total energy for hot carrier generation has to be found. Moreover, Brown and colleagues \cite{brown2016plasmondecay} also draw attention to the trade-off between introduction of efficient interband-based hot carrier generation channel and the associated weakening of the field enhancement. 

However, our results, especially for the Pd-Au system, show that absorption can actually be increased (at the expense of scattering) despite partial quenching of the field. This means that for nanostructures of appropriate size and geometry, the effects can actually act synergistically: the increase in hot carrier generation propensity can be accompanied by the simultaneous increase in absorption, creating a best-of-both-worlds situation. The same principle should also hold for absorption increase by d-band tuning for intermediate losses in Au-Ag systems (section 3.1). 

Additionally, the alloying-induced absorption increase that occurs without d-band involvement (at energies below the d-band threshold), as described in Section 3.2, could also be accompanied by the increase in hot carrier generation propensity. The emergent interband transitions, proposed as one of the causes of increased losses, are also suspected to benefit hot carrier generation \cite{bubas2024hotcarrier}. 

In light of this discussion, we estimate that the increase in hot carrier production might be greater than the increase in absorption when alloys are used. As more energy is absorbed, a greater proportion of that energy is directed to creating hot carriers due to electronic structure changes related to alloying. Multiplicative nature of this relation points to promising outsized efects on total hot carrier generation.

\textbf{Enhancing photothermal applications:}
The most straightforward way of turning the absorbed light into heat is by Ohmic losses. Additionally, hot carriers can also thermalize producing heat. Whether generated directly or by hot carrier thermalization, heating of the nanoparticle can lead to temperature increase even above 200 $^{\circ}$C \cite{biswas2023photothermally}. Absorption maximization by alloying can thus be a direct improvement for photothermal applications where pure metals already find use. Many of them are biomedical applications which greatly benefit from optical response in the near-infrared window, where Au is the standard choice. On that note, our work demonstrated that improvement up to an order of magnitude can be expected by alloying a small amount of Pd with Au in the energy range of interest. Fine tuning near-infrared absorption by Au-Ag alloying in the  can also be of great benefit. Moreover, since alloying was shown to broaden the absorption band, a potential improvement can be found in applications benefiting from full solar spectrum utilization such as solar vapor generation \cite{liang2019plasmon}. 

\textbf{Perspective for low loss alloys:}
Finding the utilization of alloys with losses lower than both constituent metals seems to be straightforward but examples of such alloys are extremely scarce. The quest for materials with optical losses that are lower than those of pure metals was long pursued in plasmonics without significant success. The attempts include low loss alloys \cite{cortie2019thequest, keast2014firstprinciples}. Despite that, in certain energy ranges, several alloys have been identified as potentially less lossy than the the good plasmonic metals that constitute them \cite{keast2014firstprinciples, keast2013aual2andptal2, blaber2009optical}. 

Here we can provide a general guideline to devise prospective plasmonic alloys less lossy than their constituent metals. The strategy relies on combining a stable d-shell metal, whose d-bands lead to optical losses in the visible spectrum, with a stable p-shell metal the d-band edge in the alloy is generally lowered in energy. Alongside the presented case of Cu-Al alloy, this effect can also be seen for AuAl$_2$ and PtAl$_2$ intermetallics \cite{shahcheraghi2016anomalously, keast2013aual2andptal2}. In this way, the energy region in which the the quenching influence of d-bands can be avoided is expanded while ensuring the stability and applicability of the alloy. Such a perspective can be especially interesting for introduction of metals such as Pt and Pd, with high d-bands, into general plasmonics-applications. Proof of this concept is demonstrated using DFT calculations of DOS, which show lowering of the d-bands of Pd and Pt by alloying with Al (Figures S7 and S8).

\section{Conclusions}
In this work we showed how, depending on a particle's size and shape, the absorption of a plasmonic nanoparticle can be maximized by tuning the optical losses. To achieve that, we first elucidated different electrodynamic effects that govern absorption. In short, the absorption efficiency is proportional to the optical losses and to the field inside the particle, which is quenched by both optical losses and radiation damping. 

The main finding is that the optical losses should be as high as possible, as long as they are smaller than radiation damping. An interesting novel finding is that dynamic depolarization mitigates the field quenching of optical losses, especially for larger and elongated particles. Combined with larger role of radiation damping in such particles, we conclude that minimal losses are desirable only for small spherical particles, while for larger and nonspherical particles higher optical losses are needed to maximize absorption. Compared to oversimplistic guidelines based on seeking minimal losses to avoid plasmon quenching, this analysis provides deeper insights and complete more nuanced guidelines for optimization of optical losses.

As the most suitable way to achieve optimal losses we highlight alloying. By combining different metals in appropriate ratios the alloy electronic structure can be deliberately tuned - even beyond the properties exhibited by pure metals. The losses that maximize absorption are sometimes lower, higher, or in between those of pure metals. For each case we presented a tuning strategy utilizing alloying. 

To achieve intermediate losses, d-band position can be shifted to desired levels, as exemplified in the visible range for Ag-Au system, and in the NIR for Au-Pd system. To achieve losses higher than both elements, emergent low-energy interband transitions of Ag-Au system were utilized between IR and visible range. Lower losses were achieved by combining elements with different electronic structures, such as p-shell Al and d-shell Cu. In all aforementioned cases alloy nanoparticles exhibited a substantially increased absorption, even up to an order of magnitude, when compared to pure metal nanoparticles with the same geometry. These results were generalized to alloys of other metals based on their electronic structure.

Finally, a thorough discussion is presented on how the obtained results can advance absorption-based applications that make use of photothermal effects and hot carrier generation, such as photocatalysis, solar energy harvesting, and cancer therapy. Additionally, guidelines for low-loss alloy construction were also discussed. 

\begin{acknowledgement}

This research acknowledge the financial support of the Croatian Science Foundation through the project IP-2019-04-5424.

\end{acknowledgement}

\begin{suppinfo}

Theoretical analysis of the absorption efficiency for small spherical and spheroidal particles, corresponding supporting simulations and density functional theory calculations of alloys. 

\end{suppinfo}

\bibliography{acs-achemso}

\end{document}


\maketitle

\section{The absorption efficiency in the Rayleigh approximation}
Let us assume a small particle with radius $a$ and complex dielectric function $\varepsilon_p = \varepsilon_r+ i \varepsilon_i$ embedded in a medium with dielectric function $\varepsilon_m$ and excited by a plane wave with wavelength $\lambda$. Under the assumption that scattering is negligible, the absorption efficiency is determined by the imaginary part of the static particle polarizability:
\begin{equation}
    Q_{abs} \approxeq 4 x  \textrm{Im}\left[{\frac{\varepsilon_p-\varepsilon_m}{\varepsilon_p+2\varepsilon_m}}\right] =12 x \frac{\varepsilon_m \varepsilon_i}{\left(\varepsilon_r+2\varepsilon_m\right)^2+\varepsilon_i^2} \:,
    \label{eq:Rayleigh}
\end{equation}
with $x = 2 \pi \sqrt{\varepsilon_m}a/\lambda$ being the particle size factor. According to this expression,  maximizing $Q_{abs}$ requires $\varepsilon_r \rightarrow -2\varepsilon_m$ and $\varepsilon_i \rightarrow 0$. Strictly speaking, this expression can not be applied to non-absorbing particles as it would violate unitarity \cite{chylek1979nonunitarity}. Nevertheless, the Rayleigh approximation is often used to point out the unwanted effect of optical losses in absorption.

In the electrostatic limit, $\textbf{E}_p =  \frac{3\varepsilon_m}{\varepsilon_p+2\varepsilon_m}\textbf{E}_{inc}$ and Equation 1 in the article leads to the Rayleigh approximation. In the limit $\varepsilon_r \rightarrow -2\varepsilon_m$, the field intensity enhancement decays as $1/\varepsilon_i^2$ and $Q_{abs}$ increases with no bound for $\varepsilon_i \rightarrow 0$

\section{Absorption efficiency of small spherical particles: Dynamic depolarization and retardation corrections}

\begin{figure}[htbp!]
\centering
\includegraphics[width=\linewidth]{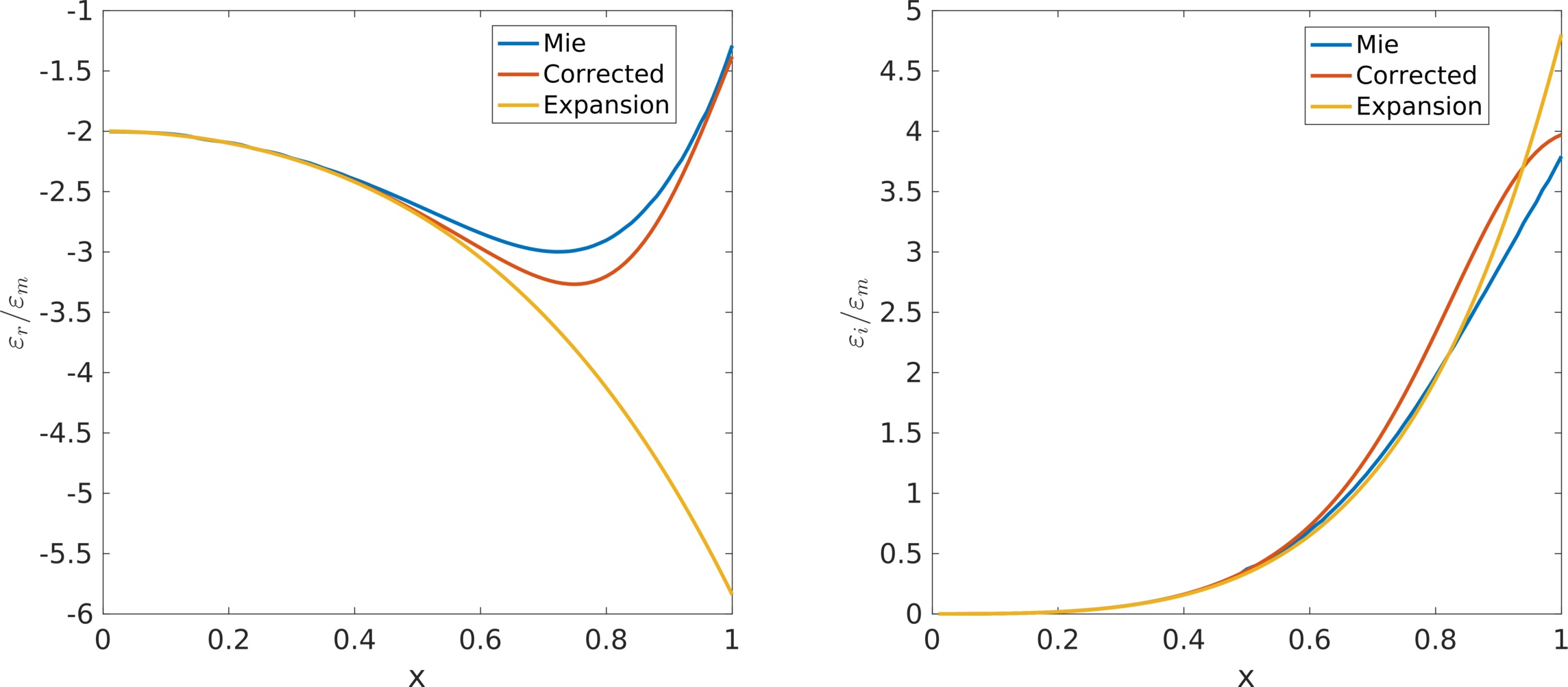}
 \caption{Optimal values of $\varepsilon_r$ (left) and $\varepsilon_i$ (right) that maximize the absorption efficiency using expressions \ref{ermax} and \ref{eimax} (red) and the series expansion (yellow). Calculations using the Mie theory and considering only the electric dipolar term $a_1$ are shown in blue.}
\label{fig:compmax}
\end{figure}

The polarization field of a particle of dielectric function $\varepsilon_p = \varepsilon_r+i\varepsilon_i$ embedded in a dielectric medium of dielectric function $\varepsilon_m$ is defined as (Gaussian units):

\begin{equation}
    \textbf{P}=\frac{1}{4 \pi}\left(\varepsilon_p-\varepsilon_m\right)\textbf{E}_{p}
\end{equation}
 with $\textbf{E}_p$ being the field inside the particle. Microscopically, the polarization can be computed from the dipole moment of the particle:

 \begin{equation}
     \textbf{P} = \varepsilon_m \frac{\alpha}{V_p} \textbf{E}_{inc}
 \end{equation}
with $\textbf{E}_{inc}$ the field incident on the particle, and $V_p$ its volume. The particle polarizability can be computed in the framework of Mie theory considering the electric dipole coefficient in the expansion of the scattered field ($a_1$, see for instance \cite{bohren2008absorption}).
\begin{equation}
    \alpha = \frac{3i}{2k^3}a_1 \approxeq \frac{\varepsilon_p-\varepsilon_m}{\varepsilon_p+2\varepsilon_m-\left(6\varepsilon_{p}-12\varepsilon_m\right)\frac{x^2}{10}-i\left(\varepsilon_{p}-\varepsilon_m\right)\frac{2x^3}{3}}r^3
    \label{alphaMie}
\end{equation}
where $k$ is the wavenumber, $r$ is the particle radius and $x$ the size factor ($x=kr$). The right-hand side is obtained by a series expansion of the Bessel and Hankel functions appearing in the Mie coefficient $a_1$ \cite{moroz2009depolarization}. Note that for $x\rightarrow 0$ one recovers the static polarizability. The terms in $x^ 2$ and $x^3$ originate from the dynamic depolarization field and the radiation damping, respectively \cite{meier1983enhanced,moroz2009depolarization}.  Inserting the polarizability in equation S2, one can express the absorption efficiency (equation 2 in the manuscript) as:
\begin{equation}
    Q_{abs} = \frac{4x}{3\varepsilon_m} \varepsilon_i \left|\frac{\textbf{E}_p}{\textbf{E}_{inc}}\right|^2 = 12 x \frac{\varepsilon_i \varepsilon_m}{\lvert \varepsilon_p+2\varepsilon_m-\left(6\varepsilon_{p}-12\varepsilon_m\right)\frac{x^2}{10}-i\left(\varepsilon_{p}-\varepsilon_m\right)\frac{2x^3}{3}\rvert^2}
\end{equation}
Maximizing this expression with respect $\varepsilon_r$ and $\varepsilon_i$ one obtains:
\begin{equation}
    \frac{\varepsilon_{r,max}}{\varepsilon_m} =  \frac{-2\left(225-81x^4-50x^6\right)}{225-270x^2+81x^4+100x^6} \approxeq -2 - \frac{12}{5} x^2 -\frac{36}{25}x^4+\mathcal{O}\left({x^6}\right) 
    \label{ermax}
\end{equation}

\begin{equation}
    \frac{\varepsilon_{i,max}}{\varepsilon_m} = \frac{90 x^3\left(x^2+5\right)}{225-270x^2+81x^4+100x^6} \approxeq 2 x^3 + \frac{14}{5} x^5+\mathcal{O}\left({x^7}\right) 
    \label{eimax}
\end{equation}
At the lowest order, Equation \ref{eimax} coincides with previous calculations \cite{tribelsky2011anomalous,tretyakov2014maximizing}. Note that the $x^3$ term in \ref{eimax} arises from radiation damping in \ref{alphaMie}, i.e. proportional to the particle volume, while the $x^5$ term results from the product of radiation damping and dynamic depolarization, as it is more explicitly shown in Section \ref{sec:spheroid}. Figure \ref{fig:compmax} compares \ref{ermax} and \ref{eimax} with the results from Mie theory,  illustrating the remarkable validity of the approximation in a wide range of $x$ values.

The maximum achievable absorption efficiency reads as:
\begin{equation}
    Q_{abs,max} = \frac{15}{2x^2\left(5+x^2\right)}.  
    \label{Qmax}
\end{equation}
Alternatively, one can compute the absorption efficiency from the extinction and scattering efficiencies taking into account only the electric dipole term:
\begin{equation}
    Q_{ext} = 4 x \textrm{Im}\left[\frac{\alpha}{r^3}\right]  
\end{equation}
\begin{equation}
    Q_{scat} = \frac{8}{3} x^4 \left|\frac{\alpha}{r^3}\right|^2  
\end{equation}
\begin{equation}
    Q_{abs} = Q_{ext}-Q_{scat}  
\end{equation}
The absorption efficiency calculated in this way is also maximized for  \ref{ermax} and \ref{eimax}. However, in this case, the maximum value is:
\begin{equation}
    Q_{abs,max} = \frac{3}{2x^2},   
\end{equation}
which agrees with \ref{Qmax} when terms on $x^4$ are ignored. Note that in both cases, the absorption cross section ($C_{abs} = \pi a^2 Q_{abs}$) remains nearly independent of the particle radius and scales as $\lambda^2$ in the small particle limit, in agreement with \cite{tribelsky2011anomalous,tretyakov2014maximizing}. 

\section{Absorption efficiency in spheroidal particles}
\label{sec:spheroid}

First we consider a prolate spheroidal particle with $a$ and $c$ being the major and (two) minor semiaxis ($c<a$). The static polarizability of the spheroid along the $j$ axis is \cite{bohren2008absorption} (Gaussian units):
\begin{equation}
\alpha_j = \frac{V_p}{4\pi}\frac{\varepsilon_p-\varepsilon_m}{\varepsilon_m+L_j (\varepsilon_p-\varepsilon_m)}    
\end{equation}
with $L_j$ the static depolarization factor at the $j$ axis, with explicit expressions given elsewhere \cite{januar2020role, bohren2008absorption} . In order to include finite-size effects in the electrodynamic response of the spheroid as it was done in the previous section, we can define an effective depolarization factor as:
\begin{equation}
L_{eff,j} = L_j-\frac{D_j V_p}{4 \pi c}k^2-i \frac{V_p}{6 \pi}k^3
\label{eqLeff}
\end{equation}
where $k$ is the wavenumber ($k=2\pi\sqrt{\varepsilon_m}/\lambda$, with $\lambda$ being the vacuum wavelength radiation), $D_j$ the dynamic depolarization factor along the $j$-axis \cite{moroz2009depolarization} and $i$ is the imaginary unity. Assuming that the field is applied along the spheroidal semimajor axis ($L\equiv L_{j\|c}$ and $D \equiv D_j$) the effective depolarization factor can be expressed as:
\begin{equation}
L_{eff} = L-\frac{D}{3}R^{2/3}x^2_{eff}-i \frac{2}{9}{x_{eff}^3}
\label{eqLeff2}
\end{equation}
where $x_{eff}$ is an effective size factor ($x_{eff} = k (a^2c)^{1/3}$) and $R=c/a$. The effective spheroid polarizability is then:
\begin{equation}
\alpha_{eff} = \frac{V_p}{4 \pi}\frac{\varepsilon_p-\varepsilon_m}{\varepsilon_m+L_{eff} (\varepsilon_p-\varepsilon_m)}=\frac{V_p}{4\pi} \tilde{\alpha}_{eff}    
\end{equation}
where $\tilde{\alpha}_{eff}$ is an effective polarizability per unit of volume. Now, by computing the absorption efficiency either in an analogous manner as done in the previous section or by the difference between extinction and scattering efficiencies, given by 
\begin{equation}
Q_{ext} = \frac{C_{ext}}{\pi x_{eff}^2}=\frac{4}{3} x_{eff}  \textrm{Im}\left[\tilde{\alpha}_{eff} \right] 
\end{equation}

\begin{equation}
Q_{scatt} = \frac{C_{scatt}}{\pi x_{eff}^2}= \frac{8}{27} x_{eff}^4 \lvert \tilde{\alpha}_{eff}\rvert^2  
\end{equation}

\begin{equation}
Q_{abs} = Q_{ext}-Q_{scatt},
\end{equation}
it can be obtained that the maximum $Q_{abs}$ is achieved for:
and 
\begin{equation}
\begin{aligned}
    \frac{\varepsilon_{r,max}}{\varepsilon_m} & = 1- \frac{81 L-27 D R^{2/3} x_{eff}^2}{81 L^2-54 D L R^{2/3} x_{eff}^2+9 D^2R^{4/3}x_{eff}^4+4x^6} \\ & \approxeq 1 - \frac{1}{L} - \frac{DR^{2/3}}{3L^2}x_{eff}^2 + \frac{D^2R^{4/3}}{9L^3}x_{eff}^4+\mathcal{O}\left({x_{eff}^6}\right)         
\end{aligned}
\label{ermaxspheroid}
\end{equation}

\begin{equation}
\begin{aligned}
    \frac{\varepsilon_{i,max}}{\varepsilon_m} & = \frac{18 x_{eff}^3}{81 L^2-54 D L R^{2/3} x_{eff}^2+9 D^2R^{4/3}x_{eff}^4+4x_{eff}^6} \\ & \approxeq  \frac{2}{9L^2}x_{eff}^3 + \frac{4 D R^{2/3}}{27L^3}x_{eff}^5+\mathcal{O}\left({x_{eff}^7}\right) 
\end{aligned}
\label{eimaxspheroid}
\end{equation}
Note that for a spherical particle ($L=1/3$, $D=1$) one does not recover expressions \ref{ermax} and \ref{eimax} because the effective depolarization factor as expressed in \ref{eqLeff} can not be recasted into the denominator of the polarizability deduced by the Mie theory (expression \ref{alphaMie}).

\begin{figure}[htbp!]
\centering
\includegraphics[width=0.9\linewidth]{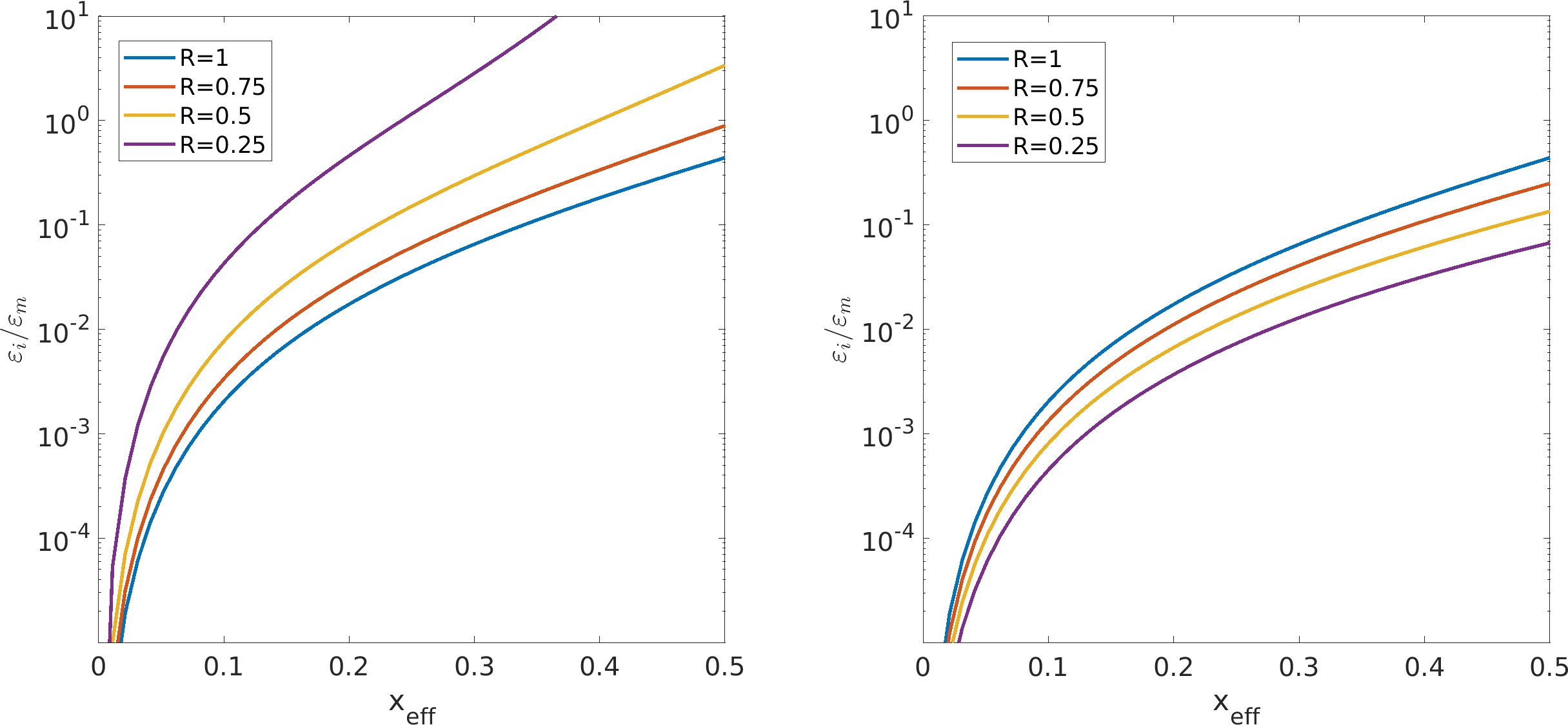}
 \caption{Optimal values of $\varepsilon_i$ that maximize the absorption efficiency of a prolate (left) and oblate (right) spheroid polarized along the symmetry semi-axis for different aspect ratios using expression \ref{eimaxspheroid}}
\label{fig:compmaxspheroid}
\end{figure}

The case of an oblate spheroid with (one) minor semi-axis $c$ and (two) major semi-axis $a$ with applied field polarized along the short, i.e. symmetry, semi-axis is very similar. Defining again $R=c/a$, the effective depolarization factor reads as:
\begin{equation}
L_{eff} = L-\frac{D}{3}R^{-2/3}x^2_{eff}-i \frac{2}{9}{x_{eff}^3}
\label{eqLeffoblate}
\end{equation}
and the real and imaginary part of the dielectric function that maximize absorption is the same as \ref{ermaxspheroid} and \ref{eimaxspheroid} with the replacements $R^{2/3} \rightarrow R^{-2/3}$ and $R^{4/3} \rightarrow R^{-4/3}$, noting that the static and dynamic depolarization factors have different expressions for prolate or oblate particles \cite{moroz2009depolarization}. Thus, in the case of the electric field polarized along the symmetry axis, the static depolarization factor decreases and the dynamic depolarization factor increases with $R$ for a prolate particle, while the opposite is true for an oblate particle. According to the previous expressions and for small $x_{eff}$ values, a prolate (oblate) spheroid will require higher (lower) losses to maximize absorption in comparison to a sphere, when polarized along the symmetry axis, as shown in Figure \ref{fig:compmaxspheroid}. When the particles are not polarized along the non-symmetry axis, the trends for the static and dynamic depolarization factors reverse and so it happens for the losses that maximize the absorption efficiency. 

\begin{figure}[htbp!]
\centering
\includegraphics[width=0.3\linewidth]{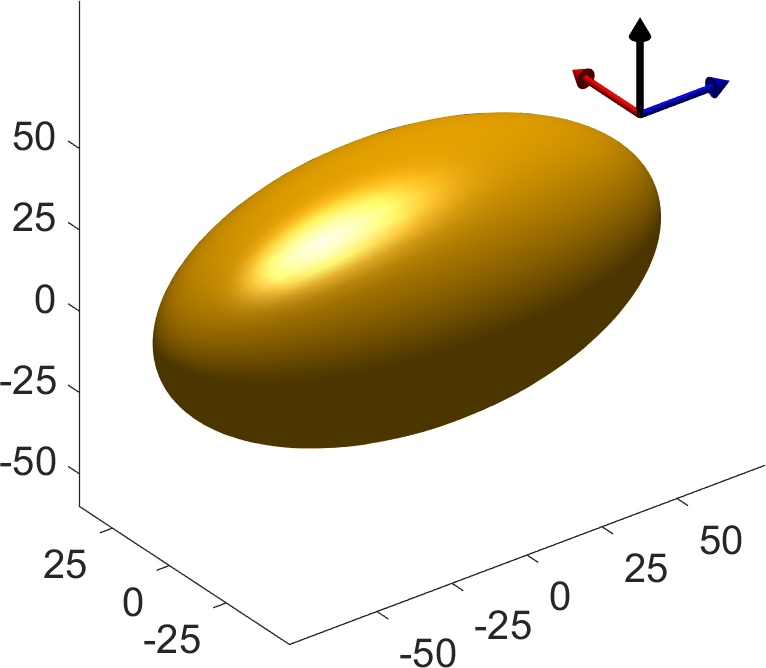}
\hspace{0.03\linewidth}
\includegraphics[width=0.3\linewidth]{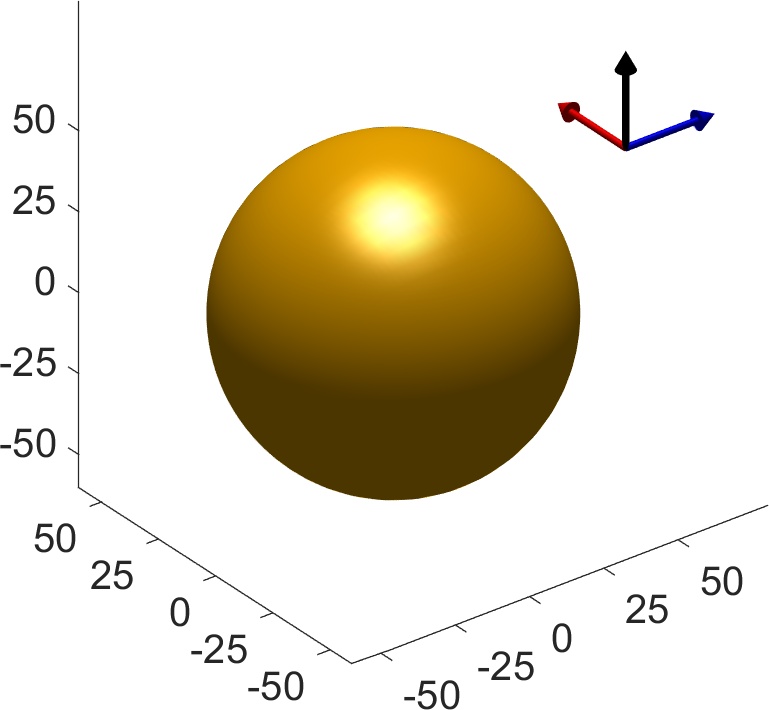}
\hspace{0.03\linewidth}
\includegraphics[width=0.3\linewidth]{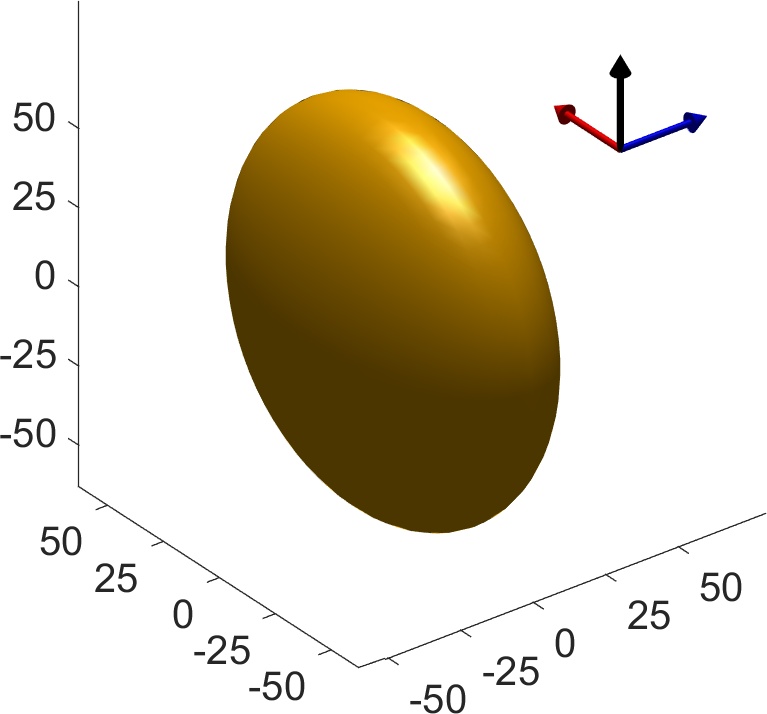}
\vspace{1cm}

\includegraphics[width=0.3\linewidth]{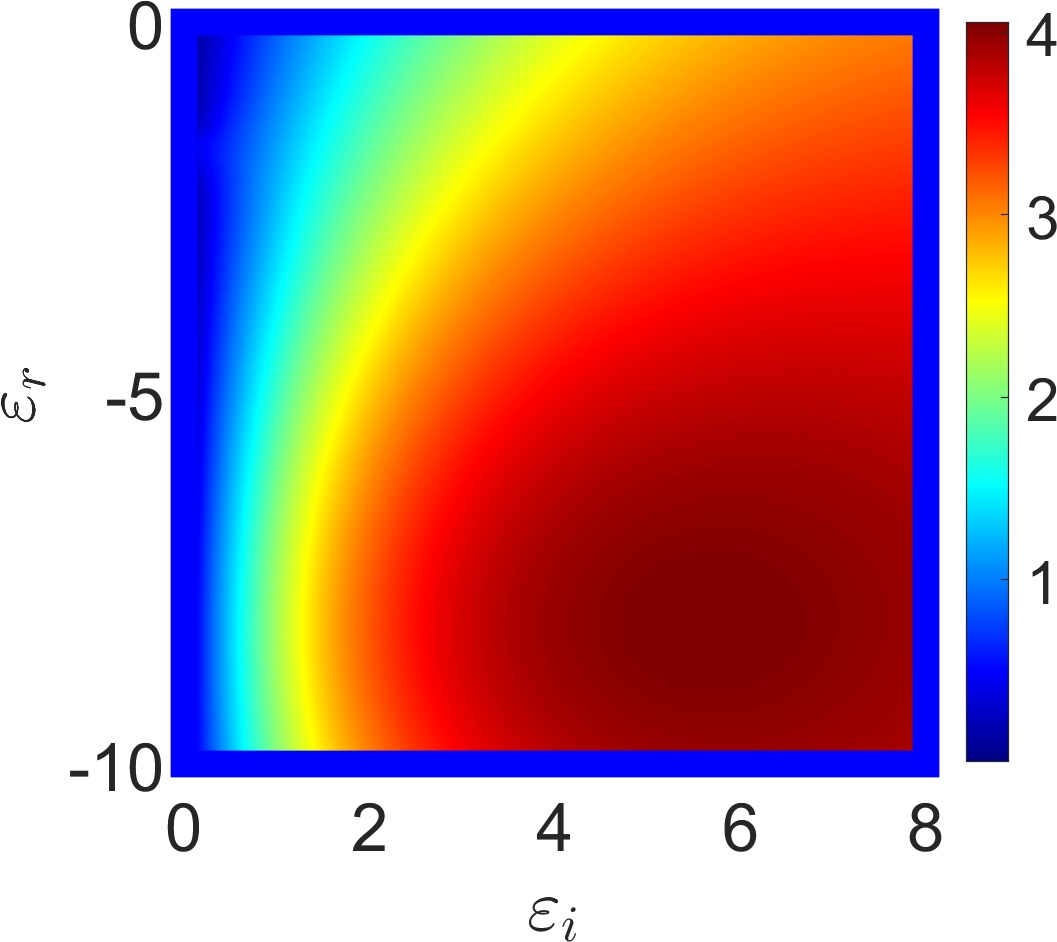}
\hspace{0.03\linewidth}
\includegraphics[width=0.3\linewidth]{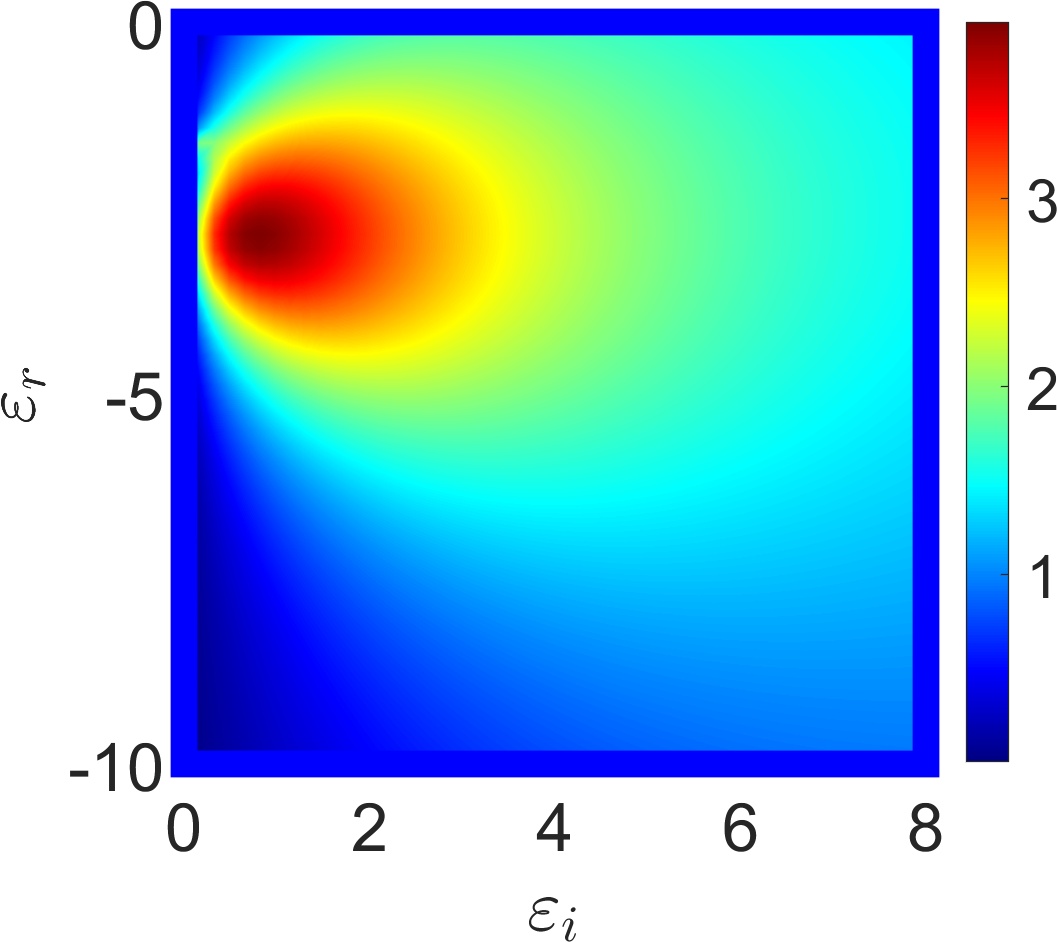}
\hspace{0.03\linewidth}
\includegraphics[width=0.3\linewidth]{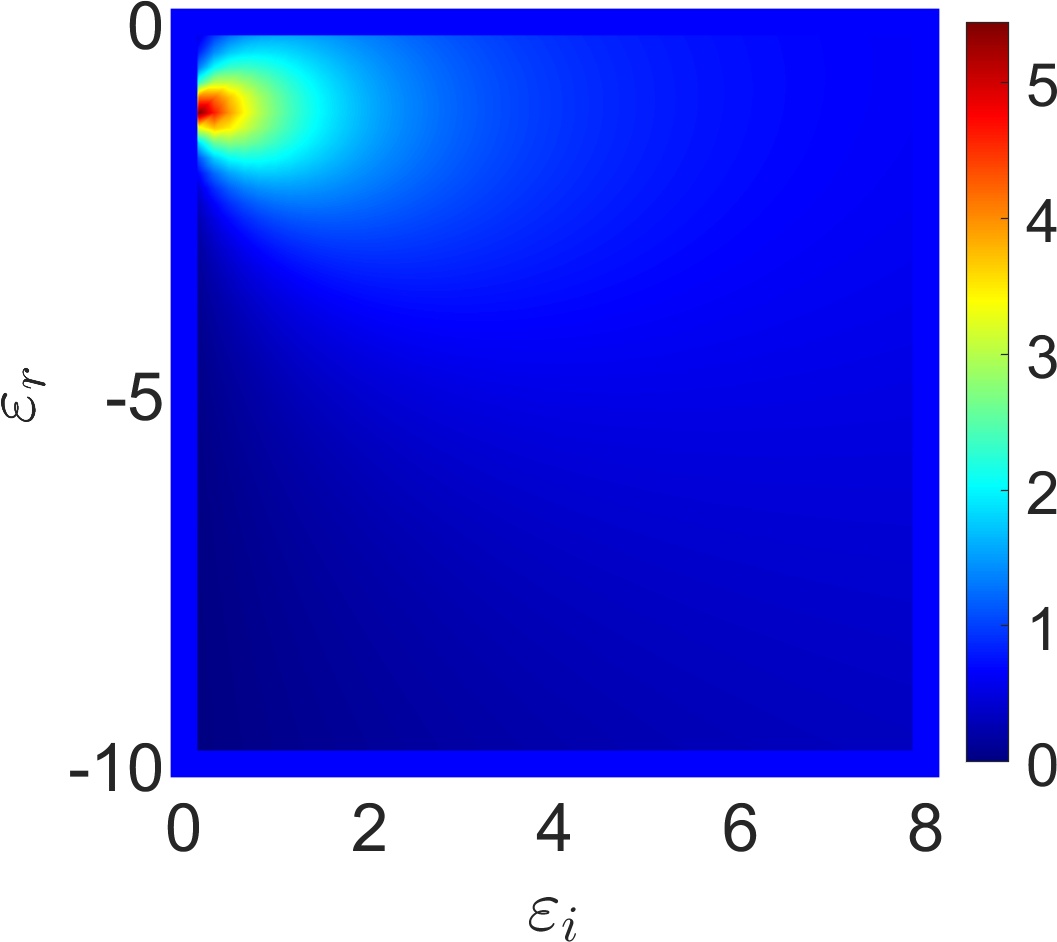}
\vspace{1cm}
 
 \includegraphics[width=0.3\linewidth]{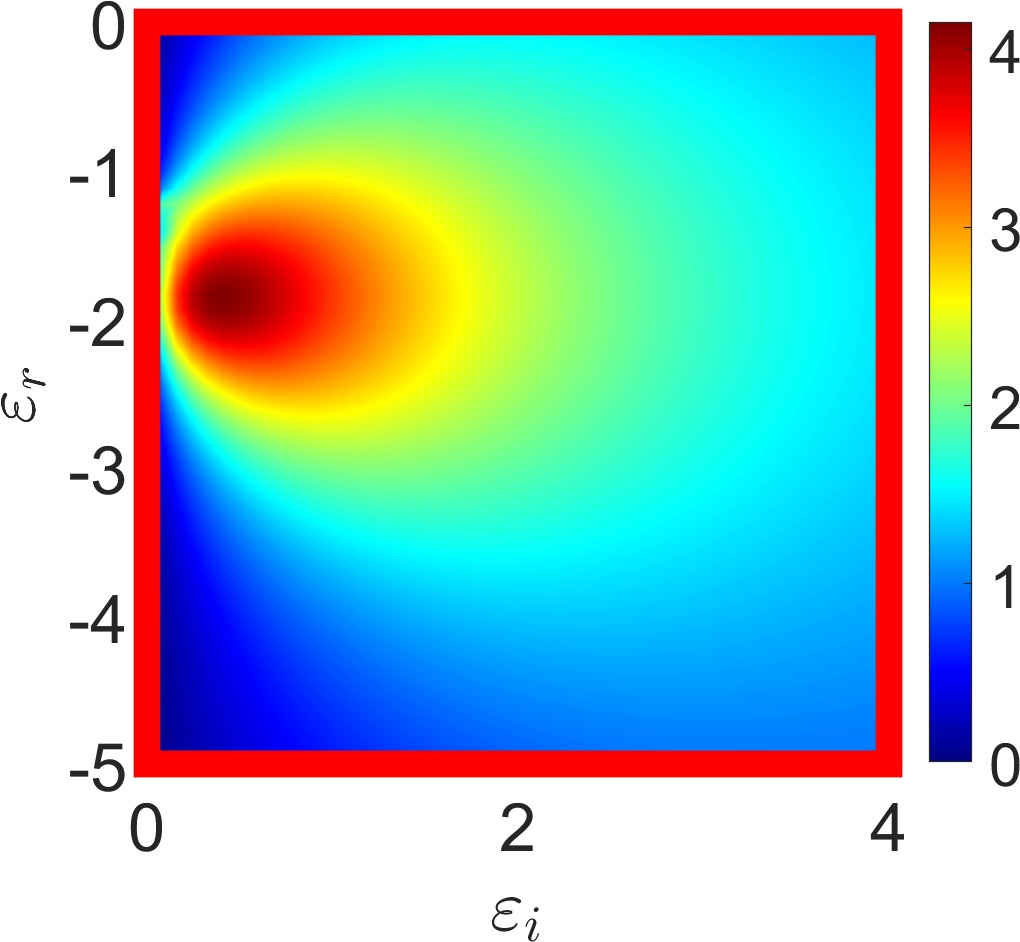}
\hspace{0.03\linewidth}
\includegraphics[width=0.3\linewidth]{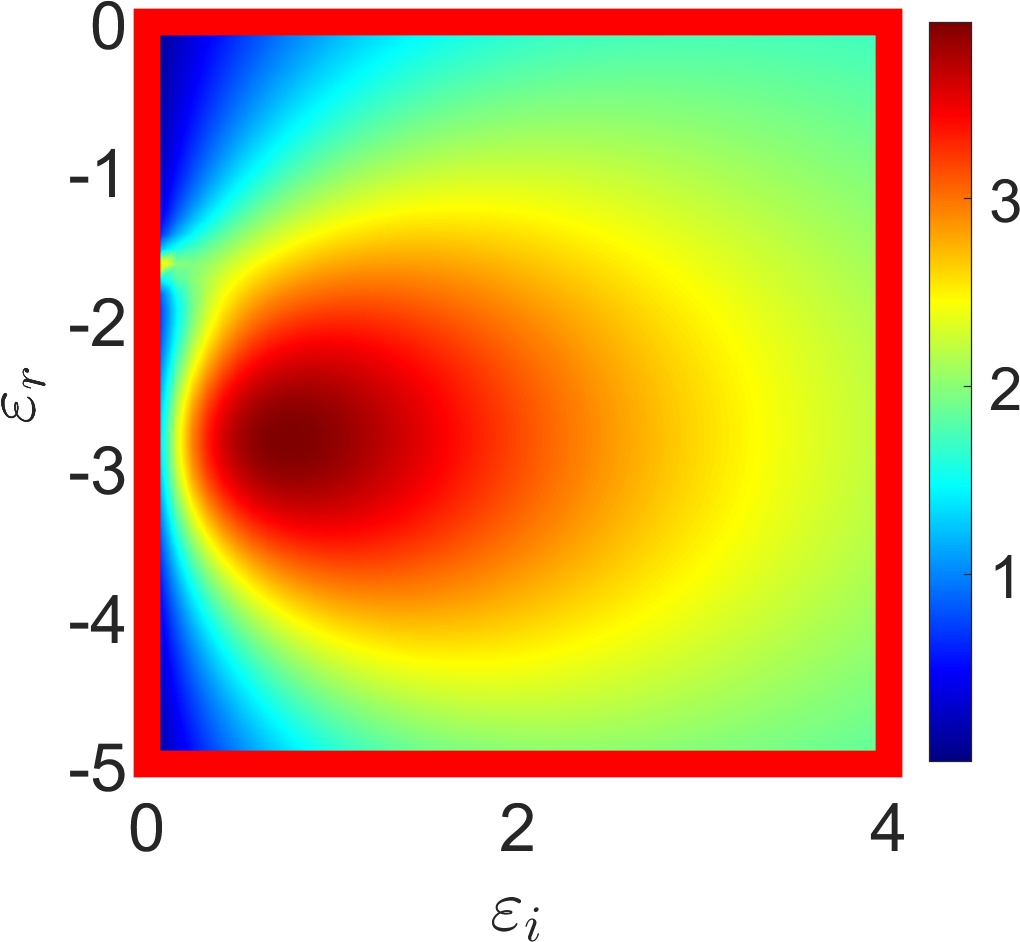}
\hspace{0.03\linewidth}
\includegraphics[width=0.3\linewidth]{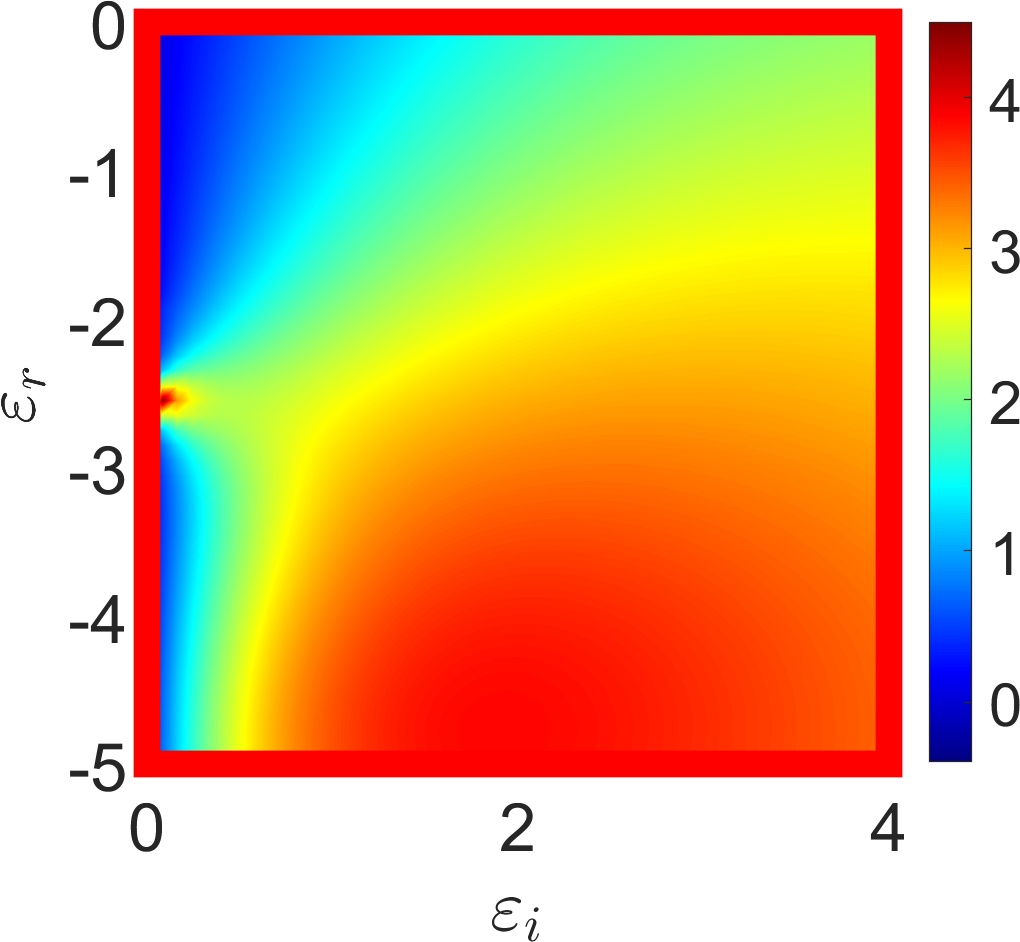}

\caption{Absorption efficiency (middle and bottom rows) as a function of the real and imaginary part of spheroidal particles (top row). All particles are supposed to have the same effective radius (25 nm). Regardless of the particle symmetry, polarization along the long (short) semi-axis results in maximum absorption efficiency take place at larger (smaller) $\varepsilon_i$ than for a spherical particle.}
\label{fig:compmaxspheroidabs}
\end{figure}

\begin{figure}[htbp!]
\centering
\includegraphics[width=\linewidth]{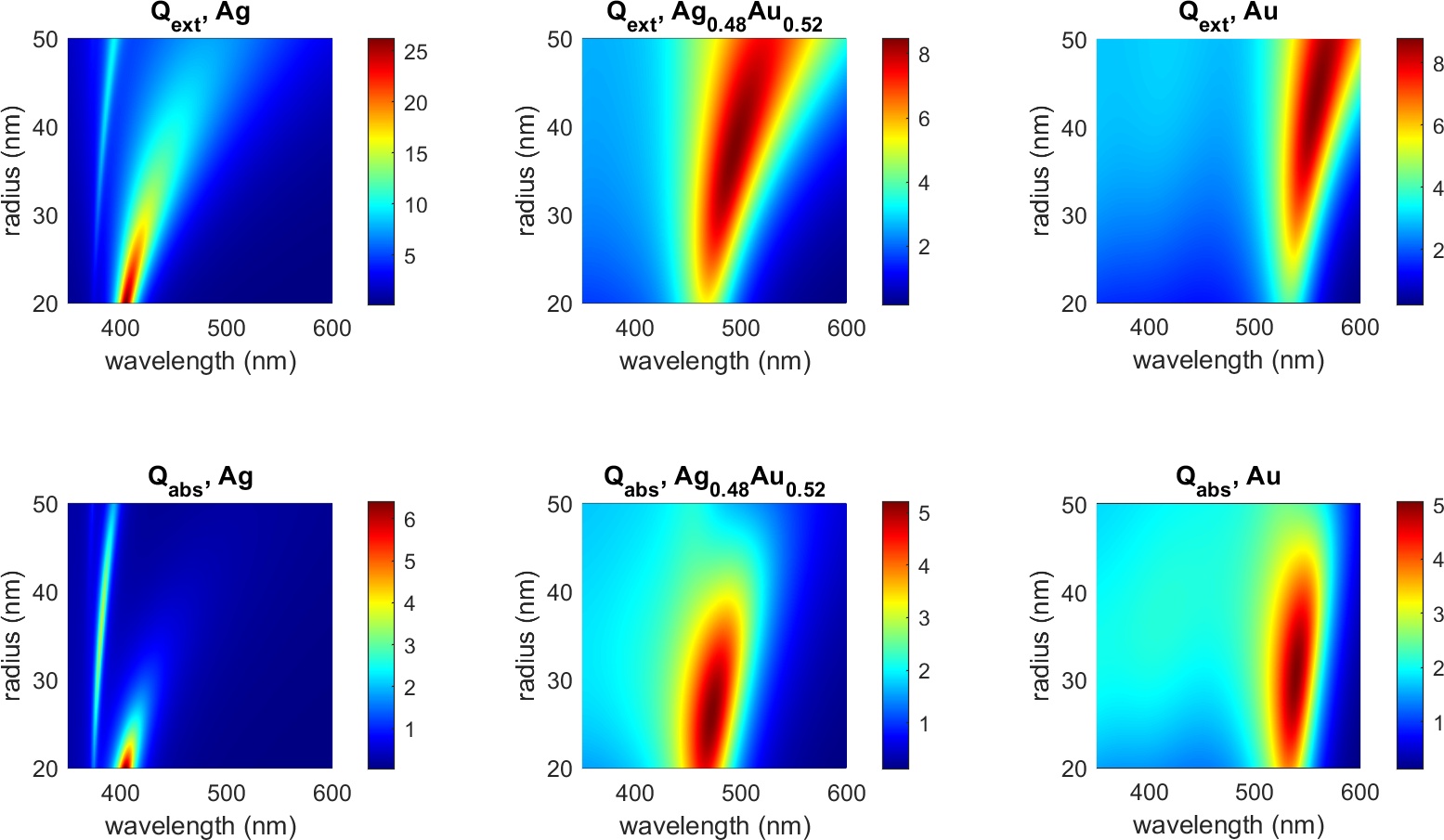}
 \caption{Extinction (top row) and absorption (bottom row) efficiencies as a function of particle radius and wavelength for Ag (left), Ag$_{0.48}$Au$_{0.52}$ (middle) and Au (right). }
\label{fig:qextqabsAgAu}
\end{figure}


\begin{figure}[htbp!]
\centering
\includegraphics[width=\linewidth]{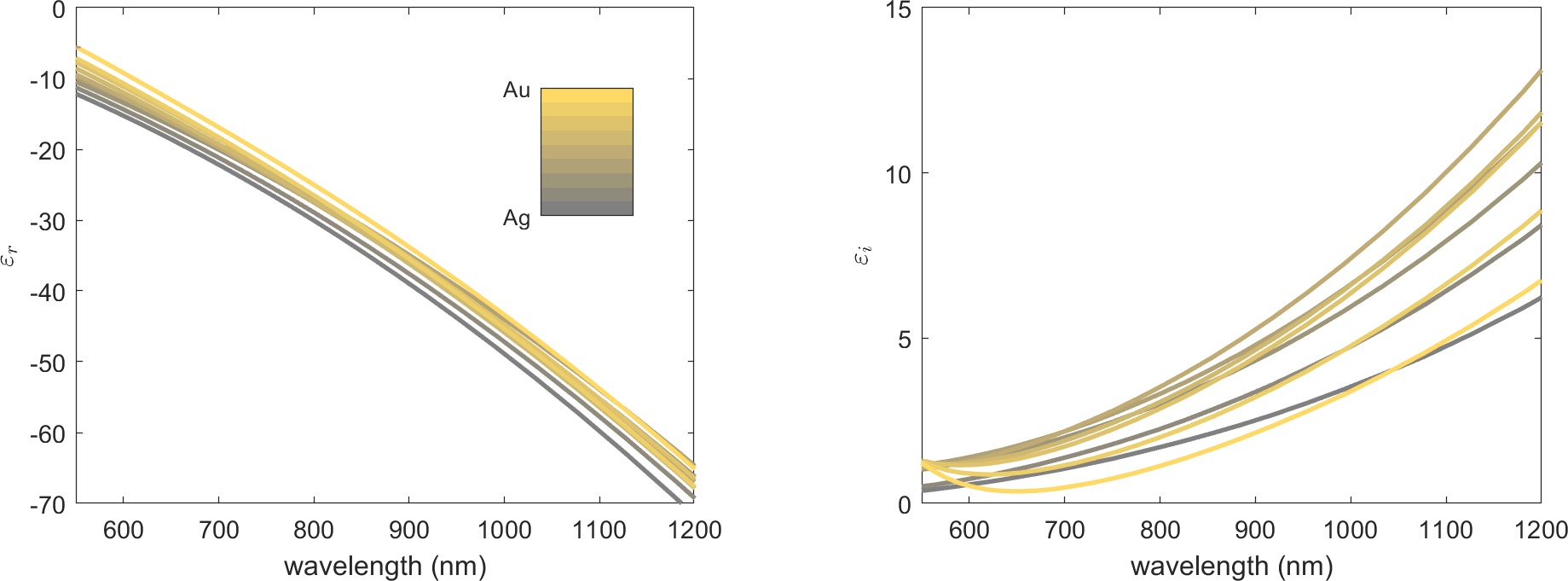}
 \caption{Real (left) and imaginary (right) part of the dielectric function of AgAu alloys taken from \cite{pena2014optical}.}
\label{fig:AgAuIR}
\end{figure}

\bibliography{sample}
